\documentclass[11pt]{article}

\usepackage{color,graphicx}
\usepackage{young}
\usepackage[vcentermath]{youngtab}
\usepackage{amsmath,amssymb,graphicx}
\usepackage{hyperref}
\definecolor{darkred}{rgb}{0.65,0.15,0}
\hypersetup{pdfborder={0 0 0},colorlinks=true,urlcolor=darkred,citecolor=blue,linkcolor=darkred,linktocpage=true}

\usepackage{cite}
\usepackage{amsmath}
\usepackage{amsfonts}
\usepackage{amssymb}
\usepackage{graphicx}%
\usepackage{amsthm}
\usepackage{mathrsfs}
\usepackage[T1]{fontenc}
\usepackage{enumerate}
\usepackage{color}

\def\4diml{four-dimensional}

\def\-1{^{-1}}

\makeatletter

\@addtoreset{equation}{section}
\makeatother

\begin{document}

\thispagestyle{empty}

\vspace{5mm}

\begin{center}
{\LARGE \bf Exact FLRW cosmological solutions via invariants\\[2mm]  of the symmetry groups }

\vspace{15mm}
\normalsize
{\large   E. Ahmadi Azar\footnote{e.ahmadi.azar@azaruniv.ac.ir},
K. Atazadeh\footnote{atazadeh@azaruniv.ac.ir}, A. Eghbali\footnote{eghbali978@gmail.com},}

\vspace{2mm}
{\small \em Department of Physics, Faculty of Basic Sciences,\\
Azarbaijan Shahid Madani University, 53714-161, Tabriz, Iran}\\

\vspace{7mm}


\vspace{6mm}

\begin{tabular}{p{15cm}}
{\small
Until now, various methods have been demonstrated to solve the Friedmann-Lama\'{\i}tre-Robertson-Walker (FLRW) equations in the
spatially flat $(k=0)$ cosmological model.
In this study, in order to solve the field equations of the spatially flat FLRW cosmological model
in the presence of $\Lambda$, a new method based on the invariants of the symmetry groups which we called ISG-method, is presented. This method is based on the extended Prelle-Singer (PS) method and it uses the Lie point symmetry, $\lambda$-symmetry and Darboux polynomials (DPs).
 We employ this method to extract systematically the two independent first integrals (or invariants) such as $I_1 (a,\dot{a})=c_1$ and $I_2 (t,a,\dot{a})=c_2$ associated to the group of the Lie point transformations keeping the Friedmann-Einstein dynamical equation (DE),  $\ddot{a}=\phi(t,a,\dot{a})$, invariant.
The obtained solutions from solving the DEs of the FLRW cosmological model by the ISG-method are explicitly written, so that they are suitable for cosmological applications.
Finally, as an application of the solutions we look at the age of universe in the presence of cosmological constant where the dominant matter of the universe is considered to be fluid with the state parameter $w$.
In this regard, we  calculate the age of universe when the dominant matter is dust.}

\end{tabular}
\vspace{-1mm}
\end{center}


\setcounter{page}{1}
\newpage
\tableofcontents

 \vspace{5mm}

\section{\label{Sec.I} Introduction}
Sophus Lie was the founder of a mathematics based on symmetry, which was also used in physics a few decades later.
In the late 19th century, he used an interesting statement about ``symmetry'' from the German mathematician
Hermann Weyl, which was also expressed by Morez in \cite{Morez}, to solve differential equations.
The use of symmetry in classical mechanics by the German mathematician, Emmy Noether, was a clear example of this issue in physics.
The results of Lie and Noether's works showed that the symmetry can help physicists in two ways:
Firstly, symmetry is a criterion for investigating the correctness of what is assumed for the laws of nature.
A law of nature can be considered correct that is consistent with the symmetry.
Secondly, in some cases, this system of direct discovery obtains new laws,
that is, the discovery of connections between phenomena that was not before.
Also, an example of the use of symmetry as a criterion for the correctness of
a law of nature, the theory of special relativity can be mentioned.
The use of symmetry is due to order and laws that are independent of unnecessary conditions.
For example, the reproducibility of experiments in different points of space and
time depends on the invariance of the laws of nature under space translation
and rotation (space homogeneity and isotropy) and time translation (time homogeneity).
Without such rules, physical events cannot be well understood and laws describing these events cannot be formulated.
The concept of symmetry and its role in the formulation of physical theories,
including gravity and cosmology, have been of interest to scientists and philosophers
through the centuries.
Newton \cite{Oliveri} established the theory of classical mechanics by introducing the laws of motion.
After Newton, Dutch mathematician, C. Huygens, used the concept of symmetry in one of his works in classical mechanics \cite{Schiffer}.
He was able to obtain the law of conservation of linear momentum by using the invariance principle.
Finally, Einstein extend the Galilean relativity principle of classical mechanics to all the laws of physics by considering the concept of symmetry;
the principle is the base of the theory of general relativity.

As mentioned earlier, a significant development in symmetry began with Lie's work in the 19th century.
Since 1873, Lie began to a systematic study of the continuous symmetry groups.
These groups keep the differential equations invariant.
The motivation for doing this work was to answer that he had clearly stated in his article in 1842 (see \cite{Stubhaug1}).
The field of the continuous groups of transformations that keep the differential equations invariant is now called ``the symmetry analysis of differential equations''.
In this way, Lie was able to establish a general theory for the integration of the ordinary differential equations (ODEs) based on their symmetry under the continuous groups of transformations.
In doing this, he was inspired by lectures of Sylov given on Galois-Abel theory.
The works of Lie and other mathematicians such as Birkhoff \cite{Birkhoff1}, Sedov \cite{Sedov1}, Ovsiannikov \cite{Ovsiannikov1},
Stephani  \cite{Stephani1} and Ibragimov \cite{Ibragimov1}
in the field of the continuous group of transformations for solving the differential equations during the last two centuries have influenced the progress of mathematics and physics especially cosmology.
The theory of Lie groups of transformations has found its way into many brunches of physics such as quantum mechanics, elementary particles, string theory, general relativity theory and the other theories of gravity. Nowadays, Lie groups of transformations which leaves the differential equations invariant, are called the one-parameter (in general multi-parameter) Lie groups of transformations. If a differential equation is invariant under the group of transformations, then it is said that this group is the symmetry group of that differential equation. In other words, the differential equation admits the group of transformations as a symmetry group \cite{Bluman1}.
In this regards, then there is a conservation law corresponding to that group. This important result was first realized by Noether, who helped Einstein in formulating the general theory of relativity. This result is known as Noether's first theorem that relates one conserved quantity to each Noether symmetry.
The successes of the theory of the invariants and Lie's symmetry ideas encouraged cosmologists to use symmetric tools to solve the DEs in gravitation and cosmology. Recently, symmetry has become one of the most important mathematical tools of cosmologists to solve cosmological equations.
Those interested in studying symmetry in cosmology can refer to references \cite{{Wei1},{Wei2},{Darabi1},{Capozziello1},{Dialektonpoulos1},{Capozziello2}, {Paliathanasis1},{Darabi2}}.

A cosmological model can be considered as $n$-dimensional $(n=1, 2, \cdots)$ dynamical system (DS). In the study of these models, one of the basic problems is the integration of the DEs. To do this, symmetry is useful, because these concepts are closely related to the invariants (or first integrals) of the DS. A new approach to find invariants is often studied by the Lie symmetry, the Noether symmetry, the Hojman symmetry, the Lutzky symmetry and Mei's form invariance \cite{{Wei2},{Noether1}}. Noether symmetry is an invariance of the functional action under the infinitesimal transformation of a continuous group.
The Mei's form invariance of a DS is an invariance of the dynamical functions such as Lagrangian, non-potential generalized force, generalized constraint force and etc. under the infinitesimal transformation of a continuous group.
Lie's method involves finding continuous transformations that keep the form of the system of differential equations unchanged.
Lutzky showed that the first integrals of a Lagrangian DS can be determined even by point symmetries, which do not preserve the action \cite{Lutzky1}.
Fu and Chen further studied the non-Noether symmetries and conserved quantities of non-conservative DSs \cite{Fu.Chen1}.
Hojman presented a new conservation law without using Lagrangian and Hamiltonian functions only based on the existence of symmetries \cite{Hojman1}.
Nowadays, this method has attracted more attention from researches so that
Zhang and Chen presented a unified form of Hojman and Lutzky's conservation laws in \cite{Zhang.Chen1}.
There, it has been shown that the conserved quantity is constructed only in terms of the symmetry generating vector of the equations of motion of the dynamic system.

The main aim of this study is to provide an analytical and algorithmic method for obtaining the solutions of the DEs of
the spatially flat $(k=0)$ FLRW cosmological model.
We solve analytically the DEs of the FLRW cosmological model in the framework of the GR theory in the presence of $\Lambda$.
Furthermore, we present the ISG-method to solve the field equations of the spatially flat FLRW cosmological model.
This method is based on the extended PS method and it uses the Lie point symmetry, $\lambda$-symmetry, and DPs.
The extended PS method combining with other symmetry methods can be a powerful algorithmic method for solving non-linear differential equations in physics
such as cosmology.
In this method, we extract systematically the two independent first integrals such as $I_1 (a,\dot{a})=c_1$ and $I_2 (t,a,\dot{a})=c_2$
associated to the group of the Lie point transformations keeping the Friedmann-Einstein DE, $\ddot{a}=\phi(t,a,\dot{a})$, invariant.
In this way, the integration constants can be written in terms of the density parameters of the perfect fluids that make up the universe.

The paper is sectionalized as follows.  We begin Sec. \ref{Sec.II} with a brief discussion on
symmetry methods; then we introduce the ISG-method as an algorithmic and analytical procedure
in order to solve the non-linear DEs in physics, especially in gravity and cosmology.
In Sec. \ref{Sec.III}, we introduce in detail the ISG-method including the basic quantities of the Lie point symmetry, extended PS method,  $\lambda$-symmetry and
DP.
A short review of the Friedmann equations in the presence of $\Lambda$ is given in Sec. \ref{Sec.IV}; moreover, in this section we show that
the FLRW cosmological model in the framework of the GR theory can be imagined as the dynamics of a particle in one dimension, in
such a way that the combination of two Friedmann equations can be reduced to a non-linear second-order ODE.
The main results of this work are presented in Sec. \ref{Sec.V}. In this section, in order to
obtain the FLRW solution in the presence of $\Lambda$ we analytically solve the Friedmann equations by ISG-method.
In Sec. \ref{Sec.VI}, we first look at the age of the universe in the presence of the cosmological constant where the dominant matter of the universe is fluid with the state parameter $w$.
Then, we obtain the age of the universe when the dominant matter is dust.
Finally, conclusions are reported in Sec. \ref{Sec.VII}.
\section{\label{Sec.II} DEs and symmetry methods}

Before introducing our method, i.e., ISG-method, let us give a brief discussion on symmetry methods.
For this purpose, we consider a one-dimensional particle DS in which a one-dimensional force is applied on the particle:
\begin{equation}
{\bf F}=\phi(t,x,\dot{x})\frac{ \partial}{\partial x }, \label{1}
\end{equation}
where $\phi(t,x,\dot{x})$ stands for the component of the force. We denote this DS by $S_{1}^2$, where the superscript $2$ is the order of the DE of the particle, while  the subscript $1$ indicates the number of DEs. Under the influence of this force, the particle moves on a straight line with an acceleration $\ddot{x}$. The configuration of this DS at
every moment of the time $t$ is given by the generalized coordinate ${\bf q}=(x)$ and the governing equation of the particle motion is given by the second-order ordinary differential equation (ODE):
\begin{equation}
\ddot{x}=\phi(t,x,\dot{x}). \label{2}
\end{equation}
If any of the symmetry methods mentioned in section \ref{Sec.I} are used to solve the ODE \eqref{2},
then, the order of the differential equation is reduced by one order.
That is by using each of the symmetry methods, causes ODE \eqref{2} to be reduced to the following first integral (or invariant):
\begin{equation}
I_1 (t,x,\dot{x})=c_1, \label{3}
\end{equation}
where $I_1$ is a known function of the independent variable $t$,
the dependent variable $x$ , the derivative $x$ with respect
to $t$, $\dot{x }:=  dx/dt$, and the integration constant $c_1$.
The first-order ODE \eqref{3} is called the first integral (or the invariant)
of the ODE \eqref{2}. This equation can be viewed as an algebraic equation,
and solved it for the variable $\dot{ x }$ in terms of the variables $t, x,$ and the constant $c_1$.
Let us suppose that the solution be as follows:
\begin{eqnarray}\label{4}
\dot{x} = \psi(t, x; c_1),
\end{eqnarray}
where $\psi$ is a known function of $t, x$ and $c_1$. Next, the first-order ODE \eqref{4} has to be solved.
If the variables $t$ and $x$ in the function $\psi$ cannot be separable, then, one has to back to symmetry again.
This time, a suitable method has to be chosen for solving the first-order ODE \eqref{4}.
If this equation admits a symmetry group, then by applying this symmetry group to \eqref{4},
its reduced equation, which is a first integral in the form $\Omega(t,x; c_1, c_2)=0$, can be obtained.
In the final step, by solving the obtained first integral as an algebraic equation for the variable $x$
in terms of the variable $t$ and the integration constants $c_1$ and $c_2$, one can find the following solution:
\begin{eqnarray}\label{5}
x(t) = \chi(t, c_1, c_2),
\end{eqnarray}
where $\chi$ is a known function of the time $t$ and integration
constants $c_1$ and $c_2$. But, if the variables $t$ and $x$ of the function $\psi$
in \eqref{4} are separable, then without needing to symmetry, firstly
one can separate the variable, then it can be solved by integration:
\begin{eqnarray}\label{6}
\frac{dx}{K_1(x)}= K_2(t; c_1)~dt,
\end{eqnarray}
where $K_1 (x)$ and $K_2 (t; c_1)$ are the separated functions of the $\psi(t, x; c_1)$, that is,
\begin{eqnarray}\label{7}
\psi(t, x; c_1):= K_1 (x) ~K_2 (t; c_1).
\end{eqnarray}
By integrating in both sides of the first-order \eqref{6}, one gets
\begin{eqnarray}\label{8}
\int \frac{dx}{K_1(x)}= \int K_2(t; c_1)~dt \Rightarrow G_1 (x)=G_2 (t; c_1, c_2 ).
\end{eqnarray}
This equation can be written as
\begin{eqnarray}\label{9}
\Xi(t, x; c_1, c_2 )=0.
\end{eqnarray}
Now, by solving the algebraic equation \eqref{9}
for the variable $x$ in terms of $t, c_1$ and $c_2$, one gets \eqref{5} which is the
general solution of the ODE \eqref{2}. In this way, by obtaining $x$ as
a function of the time $t$, the dynamics of the particle is known.

In most DSs, the reduced first-order differential equation \eqref{3} is not solvable
by the method of the separable variables. On the other hand, the differential equation
does not admit any symmetry group to help us to solve it by one
of the symmetry methods. What has to be done in such cases? To answer this
question, we have to invert a new method that does not have the weaknesses of
the previous methods and we can use it in such DSs. This new method is called ISG-method.

The ISG-method is an algorithmic and analytical procedure that has not been so far used in the
fields of gravity and cosmology. This method is a combination of the extended
PS procedure with Lie point symmetry method, $\lambda$-symmetry method and
DPs method. The main work of this method is the systematic
and algorithmic derivation of the independent invariants such as
\begin{eqnarray}\label{10}
I_1 (t, x, \dot{x},...,x^{(n-1)})=c_1, ~\cdots,~ I_n (t, x, \dot{x},...,x^{(n-1)})=c_n,
\end{eqnarray}
where $n$ is the order of the given DE:
\begin{eqnarray}\label{11}
x^{(n)}=\phi(t, x, \dot{x}, x^{(n-1)}).
\end{eqnarray}
The invariants \eqref{10} corresponding to the one-parameter
Lie groups of transformations are admitted by the $n$-order ODE \eqref{11} as the symmetry groups.
In the DS, where the DE is an ODE of the second-order,
two independent invariants are required. Only one of these invariants can be obtained by applying
any of the symmetry methods mentioned in Sec. \ref{Sec.I}.
While to fully solve the problem, two invariants are needed.
In the combined method one may extract these two independent invariants.
Let us suppose that $I_1 (t, x, \dot{x})=c_1$ and $I_2 (t, x, \dot{x})=c_2$ are the independent invariants which were obtained by
the combined method on the ODE \eqref{2}.
Next, to complete the solution of the problem, we look at these two invariants like two algebraic equations.
By solving one of these, for example, invariant $I_1 (t, x, \dot{x})=c_1$ , we find the variable $\dot{x}$ in terms of the
variables $t, x$ and the integration constant $c_1$ and then substitute it into the second invariant, $I_2 (t, x, \dot{x})=c_2$.
Solving $I_1 (t, x, \dot{x})=c_1$ for the variable $\dot{x}$ we then get $\dot{x} = \psi(t, x; c_1)$.
Plugging this solution into the invariant $I_2 (t, x, \dot{x})=c_2$, one finds an algebraic equation between the variables $t$, $x$ and
the integration constants $c_1$ and $c_2$ as in algebraic equation \eqref{9}.
Finally, by solving this algebraic equation for $x$ in terms of the variable $t$ and the integration constants $c_1, c_2$,
we arrive at the solution \eqref{5} so that it is the general solution of the ODE \eqref{2}.

\section{The ISG-method based on the extended PS method}
\label{Sec.III}
The first step in the ISG-method is to find the one-parameter Lie group of transformations which is admitted by
the ODE \eqref{2}. To do this, it is necessary to study the invariance of an ODE under a group of Lie point transformations \cite{p.olver}.
As mentioned in Sec. \ref{Sec.II}, $S_{1}^2$ is a DS with one degree of freedom.
The configuration space of this DS is $\mathbb{Q}=(x)$, where $x$ is the coordinate of the particle which has subjected to a one-dimensional force
${\bf F}=\phi(t, x, \dot{x}) \partial_x$. In general, a DS has $n (n=1,2,...)$
degrees of freedom. The configuration of this DS at each moment of the time $t$ is described by
a set of the generalized coordinates such as ${\bf q}=(q^1,...,q^n )$.
Suppose that the DEs governing this DS with $n$ degrees of freedom are the following $n$-order ODEs:
\begin{eqnarray}\label{12}
H^i(t, {\bf q},  {\bf q}^{(1)},...,{\bf q}^{(n)}):={q^{(n) i}}-\phi^i (t, {\bf q},  {\bf q}^1,...,{\bf q}^{(n-1)})=0,~~~~i=1,...,n,
\end{eqnarray}
where ${\bf q}^{(k)}:=d{\bf q}^{(k-1)}/d t,~k=1,...,n$ and $\phi^i (t, {\bf q},  {\bf q}^1,...,{\bf q}^{(n-1)}),~i=1,...,n$
are the force functions of the DS under consideration.
Now, let us assume that $G$ is the one-parameter Lie point group of transformations:
\begin{eqnarray}\label{13}
&&t \rightarrow {\bar t} =T(t, {\bf q}; \epsilon),\nonumber\\
G:&&\nonumber\\
&&{\bf q} \rightarrow {\bar {\bf q}} ={\bf Q}(t, {\bf q}; \epsilon),
\end{eqnarray}
where $\epsilon$, which is an infinitesimal positive real number:
$\epsilon \in \mathbb{R}^+ =(0~,~\infty), \epsilon \ll 1$, is called the parameter of the group $G$. Here,
$T$ and $\bf Q$ are functions of the variables $t, {\bf q} =(q^1,...,q^{n})$
and the parameter $\epsilon$.  Expanding the functions $T(t, {\bf q}; \epsilon)$ and ${\bf Q}(t, {\bf q}; \epsilon)$
in Taylor series about  $\epsilon=0$ we find that
\begin{eqnarray}\label{14}
&&{\bar t} =T(t, {\bf q}; 0) + \epsilon \frac{\partial T}{\partial \epsilon}{|_{_{\epsilon =0}}} + {\cal O} (\epsilon^2),\\
G:&&\nonumber\\
&&{\bar {\bf q}} ={\bf Q}(t, {\bf q}; 0)+ \epsilon \frac{\partial {\bf Q}}{\partial \epsilon}{|_{_{\epsilon =0}}} + {\cal O} (\epsilon^2).\label{15}
\end{eqnarray}
By denoting the functions $T(t, {\bf q}; 0)$ and ${\bf Q}(t, {\bf q}; 0)$ by $t$ and $q$, respectively, and also by defining
\begin{eqnarray}\label{16}
{\tau}(t, {\bf q}) &=&\frac{\partial T(t, {\bf q}; \epsilon)}{\partial \epsilon}{|_{_{\epsilon =0}}},\\
{\xi}(t, {\bf q}) &=&  \frac{\partial {\bf Q}(t, {\bf q}; \epsilon)}{\partial \epsilon}{|_{_{\epsilon =0}}},\label{17}
\end{eqnarray}
as the infinitesimals of the group $G$ , the transformation equations \eqref{14} and \eqref{15} become
\begin{eqnarray}\label{18}
&&{\bar t} =t+\epsilon {\tau}(t, {\bf q})+ {\cal O} (\epsilon^2),\\
G:&&\nonumber\\
&&{\bar {\bf q}} ={\bf q} + \epsilon {\xi}(t, {\bf q}) + {\cal O} (\epsilon^2).\label{19}
\end{eqnarray}
In addition, we assume that $G_n$ with the transformation equations
\begin{eqnarray}\label{20}
{\bar t} &=& T(t, {\bf q}; \epsilon),\nonumber\\
{\bar {\bf q}} &=& {\bf Q}(t, {\bf q}; \epsilon),\nonumber\\
{\bar {\bf q}}^{(i)} &=& {\bf Q}^{(i)} (t, {\bf q}, {\bf q}^{(1)},..., {\bf q}^{(i)} ; \epsilon) = \frac{D {\bf Q}^{(i-1)}  (t, {\bf q}, {\bf q}^{(1)},..., {\bf q}^{(i-1)} ; \epsilon)}{D T(t, {\bf q}; \epsilon)},~~i=1,...,n,\nonumber\\
{ {\bf Q}}^{(0)} &=& {\bf Q}(t, {\bf q}; \epsilon),
\end{eqnarray}
be the $n$-th order $(n=1,2,...)$ prolongation (or extension) of the group $G$ acting on space
$(t, {\bf q}, {\bf q}^{(1)},..., {\bf q}^{(n)}) $. If $G$ leaves ODEs \eqref{12} invariant, that is, we have
\begin{eqnarray}\label{21}
H^i(\bar {t}, {\bar {\bf q}},  {\bar{\bf q}^{(1)}},...,{\bar {\bf q}^{(n)}})= H^i(t, {\bf q},  {\bf q}^{(1)},...,{\bf q}^{(n)})=0,~~~~i=1,...,n,
\end{eqnarray}
then, we say that $G$ is a symmetry group for the system of the ODEs \eqref{12}. When $G$ is a symmetry group for a system of the ODE,
it is said that this system of the ODEs $(i=1,...,n)$
admits G as a symmetry group.
Note that in Eq. \eqref{20}, $D$ is the total derivative operator with respect to the independent variable $t$, which is defined as follows:
\begin{eqnarray}\label{22}
D&:=&\frac{\partial}{\partial t}+q^{k (1)} \frac{\partial}{\partial q^k}+q^{k (2)} \frac{\partial}{\partial q^{k (1)}}+...+
q^{k (n-1)} \frac{\partial}{\partial q^{k (n-2)}}\nonumber\\
&&~~~~~~~~~~~~~~~~~~~~~~~~~~~~~~~~~+\phi^k (t, {\bf q},  {\bf q}^{(1)},...,{\bf q}^{(n-1)}) \frac{\partial}{\partial q^{k (n-1)}},
\end{eqnarray}
where $q^{k (i)}(t):=d^i q^{k} (t)/dt^i,~k=1,...,n;~i=1,...,n-1.$
Here there is a well-known theorem that helps us to obtain the one-parameter Lie group of transformations for a given ODE of any order such
that the given ODE admits a symmetry group.
\\\\
{\bf Theorem 1.} (Lie's invariance condition). {\it For a DS $S_1^n$ with one degrees of freedom,
suppose that an n-order ODE}
\begin{eqnarray}\label{23}
H(t, {q},  {q}^{(1)},...,{q}^{(n)}):= {q}^{(n)} - \phi(t, {q},  {q}^{(1)},...,{q}^{(n-1)})=0,~~~~n=1, 2, \cdots,
\end{eqnarray}
{\it be the governing (or dynamical) equation, and}
\begin{eqnarray}\label{24}
{\bf X}= \tau(t, q) \frac{\partial}{\partial t}+\xi(t, q) \frac{\partial}{\partial q},
\end{eqnarray}
{\it be the infinitesimal generator of the one-parameter Lie group of transformations}
\begin{eqnarray}\label{25}
&&{\bar t} = T(t, q; \epsilon)= t+\epsilon {\tau}(t, {q})+ {\cal O} (\epsilon^2),\nonumber\\
G:&&\nonumber\\
&&{\bar {q}} = Q(t, q; \epsilon)= {q} + \epsilon {\xi}(t, {q}) + {\cal O} (\epsilon^2),~~\epsilon \ll 1,
\end{eqnarray}
{\it acting on $(t,q)$-space. Also, suppose that}
\begin{eqnarray}\label{26}
{\bf X}^{(n)}={\tau}(t, {q}) \frac{\partial}{\partial t}+
{\xi}(t, {q}) \frac{\partial}{\partial q}+
{\xi}^{(1)}(t, {q}, q^{(1)}) \frac{\partial}{\partial q^{(1)}}+...+
{\xi}^{(n)}(t, {q}, q^{(1)},...q^{(n)}) \frac{\partial}{\partial q^{(n)}},
\end{eqnarray}
{\it be the $n$-order extended infinitesimal generator of \eqref{24}, where}
\begin{eqnarray}\label{27}
{\xi}^{(k)}(t, {q}, q^{(1)},...q^{(k)}) =
\begin{cases}
D {\xi}^{(k-1)}- q^{(k)} D {\tau}(t, {q})&~~~~~~ k=1, 2, ...~~~~~\\
{\xi}(t, {q})&~~~~~~k=0.\\
\end{cases}
\end{eqnarray}
{\it The one-parameter Lie group of transformations is admitted by ODE \eqref{23} if and only if }
\begin{eqnarray}\label{28}
{\bf X}^{(n)} H({t}, q,  q^{(1)},...,q^{(n)}){|_{_{H =0}}}=0.
\end{eqnarray}
{\it This equation is called the Lie's invariance condition which is the fundamental equation of the Lie symmetry analysis.}

Below we employ the ISG-method to solve a  non-linear second-order non-linear ODE.
Before we proceed to investigate this further, let us express the following theorem \cite{Duarte1} which is presented the PS determining equations.
\\
{\bf Theorem 2.}  {\it Let $\ddot{x} = \phi(t, x, \dot{x})$ be a non-linear second-order ODE, where the
force function $\phi(t, x, \dot{x})=P/Q$ is a fractional function of polynomials $P$ and $Q$ of the variables $t, x$, and $\dot{x}$
with coefficients in the set of complex numbers. If this ODE admits a first integral such as $I(t, x, \dot{x})=c$,
then\footnote{ The equation \eqref{29} is called the ``Duarte's integral formula''.}}
\begin{eqnarray}\label{29}
I(t, x, \dot{x}) = r_1+ r_2- \int \Big[R + \frac{\partial}{\partial \dot{x}}( r_1+ r_2) \Big] d \dot{x},
\end{eqnarray}
{\it where}
\begin{eqnarray}\label{30}
r_1 = \int R(\phi + S \dot{x})  d t,~~~~~r_2 =- \int \Big(R S + \frac{\partial r_1}{\partial x} \Big)  d x.
\end{eqnarray}
{\it In Eqs. \eqref{29} and \eqref{30}, $S$ and $R$, which satisfy in the Prelle-Singer determining equations: }
\begin{eqnarray}
D[S]&=&-\phi_{_x} + S \phi_{_{\dot{x}}} +S^2,\label{31}\\
D[R]&=&-R(S + \phi_{_{\dot{x}}}),\label{32}\\
R_x&=& R_{_{\dot{x}}} S + R  S_{_{\dot{x}}},\label{33}
\end{eqnarray}
{\it  are called the null form and the integrating factor of the DE, $\ddot{x}=\phi(t, x, \dot{x})$, respectively.
Here, $\phi_{_x}:=\frac{\partial \phi}{\partial x}, \phi_{_{\dot{x}}}:=\frac{\partial \phi}{\partial \dot{x}}$ and so on.}\\


As mentioned above we shall solve a non-linear second-order ODE by making use of the ISG-method.
We fulfill this work in the following seven steps.
\\
{\bf {\small Step 1:}}~We reconsider the DS with one degree of freedom, $S_{1}^2$.
As mentioned in Sec. \ref{Sec.II}, the governing equation of this system is $\ddot{x}=\phi(t, x, \dot{x})$, where $\phi(t, x, \dot{x})$
is the force function of a particle moving in a 1-dimensional mini-super space with configuration $\mathbb{Q}=(x)$,
where $x$ is the coordinate of the particle.\\
{\bf {\small Step 2:}}~ An one-parameter Lie group of point transformations $G$ with transformation equations
\begin{eqnarray}\label{34}
&&{ t} \rightarrow {\bar t} =  t+\epsilon {\tau}(t, x)+ {\cal O} (\epsilon^2),\nonumber\\
G:&&\nonumber\\
&&x \rightarrow {\bar x} =  {x} + \epsilon {\xi}(t, {x}) + {\cal O} (\epsilon^2),~~\epsilon \ll 1,
\end{eqnarray}
must be found such that the ODE $\ddot{x} =\phi(t, x, \dot{x})$ remains invariant under this group of transformations, i.e.
\begin{eqnarray}\label{35}
\ddot{x}-\phi(t, x, \dot{x}) = \ddot{\bar x}-\phi({\bar t}, {\bar x}, \dot{{\bar x}}) =0.
\end{eqnarray}
For this purpose, by using the criterion theorem for the invariance of an ODE (Theorem 1), a set of the PDEs should be extracted for the
infinitesimals $\tau(t,x)$ and ${\xi}(t,x)$ of group $G$. By simultaneously solving this system of PDEs,
the infinitesimals $\tau$ and $\xi$ as functions of $t$, and $x$ should be obtained.\\
{\bf {\small Step 3:}}~ All Lie point symmetry vectors of the transformations group $G$ must be formed by using the infinitesimals
$\tau(t,x)$ and  $\xi(t,x)$ as ${\bf X}=\tau(t,x) \partial_{_t}+\xi(t,x) \partial_{_x}$. \\
{\bf {\small Step 4:}}~ Using the extended PS method, the determining equations \eqref{31}-\eqref{33} for the  basic quantities, the null form $S$ and the integrating factor $R$ corresponding to the DE
$\ddot{x} =\phi(t, x, \dot{x})$, should be considered.
To calculate the first integral \eqref{29}, the S and $R$ should be obtained from solving Eqs. \eqref{31} and \eqref{32}.
It is difficult to solve these simultaneously, except in special simple cases.
To obtain these two basic functions in the extended PS method, we must resort to other symmetry methods.
The most important other symmetry methods that can be used here are: (a) Lie point symmetry, (b) $\lambda$-symmetry,
(c) DPs. Therefore, the functions $S$ and $R$ should be calculated indirectly by the symmetry methods (a)-(c).\\
{\bf {\small Step 5:}}~ One may define the characteristic $Q(t, x, \dot{x}):= \xi - \dot{x} \tau$ and $\lambda(t, x, \dot{x}):=  D[Q]/Q$ by using the Lie point symmetry \eqref{34}.
Then, it can be shown that $-D[Q]/Q$ is a solution of the determining equation \eqref{31}, and thus it follows that $S(t, x, \dot{x})=-\lambda(t, x, \dot{x})$.\\
{\bf {\small Step 6:}}~ In order to introduce the DP method we give the following definition:
\\
{\bf Definition 1.} {\it Suppose $S_{1}^2$ be a DS with DE $\ddot{x}=\phi(t, x, \dot{x})$. We say that function $F(t,x,\dot{x})$  is a DP of the DS $S_{1}^2$, if it satisfies in the following Darboux's eigenvalue equation}
\begin{eqnarray}\label{36}
D[F] =
\begin{cases}
\phi_{_{\dot{x}}} F&~~~~ $if DE is not an explicit function of$~ t,\\
K F&~~~~$if DE is an explicit function of$~ t.\\
\end{cases}
\end{eqnarray}
{\it In fact, this equation is the determining equation of the DP $F$, and the eigenvalue $K(t,x,\dot{x}) (\phi_{\dot{x}})$  is called the cofactor of $F$
when DE is an explicit function of  $t$ (when DE is not an explicit function of  $t$) and has degree at most 2 \cite{Llibre.Valls}.}\footnote{Every DP $F$ defines an invariant algebraic hypersurface $F=0$, i.e., if a trajectory of
the particle in the one-dimensional mini-super space $\Bbb{Q}=(x)$ has a point in $F=0$, then the whole trajectory of the particle is contained in the hypersurface $F=0$  \cite{J.Dumortier1}.}

According to above definition, DP $F$ and the eigenvalue $K$ should be obtained.
It can be shown that the ratio $Q/F$ is a general
solution of the determining equation \eqref{32}, and hence  $R(t, x, \dot{x})=Q/F$. In this way, the null
form $S$ and the integrating factor $R$ are obtained indirectly by the Lie point symmetry method and DP, respectively,
without the need to solve their determining equations.

The DP method was introduced by Jean-Gaston Darboux in \cite{Darboux1}.
Other titles that are used by the authors for these polynomials are special integrals, special polynomials,
algebraic invariant curves, special algebraic solutions or the special polynomials \cite{Mohanasubha1}.
The DP provides us a convenient way to find the integrating factor.
If a DS possesses a sufficient number of DPs,
then one can provide a first integral from them.
In \cite{Jouanolou1}, Jouanolou showed that if the number of DPs of a polynomial system of degrees $n$ are at
least $[(n+1)/2]+2$, then the system has a rational first integral, and consequently all its obtained
solutions are invariant algebraic curves \cite{{J.Dumortier1},{J.Chavarriga1},{Christopher1},{Llibre1},{Gareia1}}.
Singer proved that \cite{Singer1} if a polynomial DS has a Liouvillian first integral,
then it can be constructed by using the invariant algebraic curves and the exponential factors of the DS.\\
{\bf {\small Step 7:}}~ If the DE $\ddot{x}=\phi(t, x, \dot{x})$ does not have any Lie point symmetry,
we should go to its generalized symmetries. One of these well-known symmetries is $\lambda$-symmetry.
In order to find  the general solution of the non-linear second-order ODEs, $\lambda$-symmetry was introduced by Muriel and Romero in \cite{{Muriel1},{Muriel2},{Muriel3}}.
The $\lambda$-symmetry method enables us to determine the integrating factor $R$.
In order to see how this symmetry method can be used to solve ODE,
it is necessary to define $\lambda$-symmetry and explain how it can be used to calculate the integrating factor.
\\
{\bf Definition 2.} {\it Suppose the second-order ODE $H(t, x, \dot{x}, \ddot{x}):=\ddot{x}-\phi(t, x, \dot{x})=0$
is the DE of a DS.The vector field ${\bf X}=\tau(t,x) \partial_{_t}+\xi(t,x) \partial_{_x}$ is an infinitesimal generator of a
$\lambda$-symmetry of this system, if there exists a function such as $\lambda(t, x, \dot{x})$ which satisfies in the following relation:}
\begin{eqnarray}\label{37}
{{{\bf X}{^{[\lambda, (2)]}} [H(t, x, \dot{x}, \ddot{x})]}{\Big |}}_{H=0}=0,
\end{eqnarray}
{\it where}
\begin{eqnarray}\label{38}
{\bf X}{^{[\lambda, (2)]}} :=\tau \frac{\partial}{\partial t} +\xi \frac{\partial}{\partial x}+
{\xi}{^{[\lambda, (1)]}} \frac{\partial}{\partial \dot x}  + {\xi}{^{[\lambda, (2)]}} \frac{\partial}{\partial \ddot x},
\end{eqnarray}
{\it in which, the functions ${\xi}{^{[\lambda, (1)]}}$  and ${\xi}{^{[\lambda, (2)]}}$  are the first- and second-order
extensions of the vector field ${\bf X}=\tau(t,x) \partial_{_t}+\xi(t,x) \partial_{_x}$, which are defined as follows:}
\begin{eqnarray}
{\xi}{^{[\lambda, (1)]}} &:=& (D+\lambda) {\xi}{^{[\lambda, (0)]}} (t, x)-\dot x (D+\lambda)\big(\tau(t, x)\big), \nonumber\\
{\xi}{^{[\lambda, (2)]}} &:=& (D+\lambda) {\xi}{^{[\lambda, (1)]}} (t, x)-\dot x (D+\lambda)\big(\tau(t, x)\big).\label{39}
\end{eqnarray}

By substituting the extensions of \eqref{39} into the $\lambda$-symmetry condition \eqref{37} one gets
\begin{eqnarray}
{\tau} \phi_{_t} + {\xi} \phi_{_x}+{\xi}{^{[\lambda, (1)]}} \phi_{_{\dot x}} - {\xi}{^{[\lambda, (2)]}} =0.\label{40}
\end{eqnarray}
This equation can be written as $D^2 [Q]=\phi_{_{\dot x}}  D[Q]+\phi_{_x} Q$, where $Q:=\xi -\dot x \tau$.
If one sets $\lambda=0$ in Eq. \eqref{37}, then Lie's invariance condition can be obtained.
By solving the $\lambda$-symmetry condition, Eq. \eqref{37}, one can obtain the explicit forms of the functions
$\tau$, $\xi$ and $\lambda$. When these functions are known,
then the first integral can be calculated by the following processes \cite{{Mohanasubha1},{Muriel2},{Bhuvaneswari1}}:
\\
$\bullet$~We find a first integral $w(t, x, \dot x)$ associated to the symmetry vector ${\bf X}{^{[\lambda, (1)]}}$
which is the particular solution of the partial differential equation $w_{_x} + \lambda w_{_{\dot x}}=0$,
where  $w_{_x}:={\partial w}/{\partial x}$,  $w_{_{\dot x}} :={\partial w}/{\partial \dot x}$,
and  ${\bf X}{^{[\lambda, (1)]}}$ is the first-order $\lambda$-prolongation (or extension) of the vector field
${\bf X}=\tau(t,x) \partial_{_t}+\xi(t,x) \partial_{_x}$.\\
$\bullet$~ We calculate $D(w)$ and express it in terms of $(t, w)$, i.e., in the form $D(w)= F(t, w)$.\\
$\bullet$~ We find a first integral $G$ of $\partial_t +F(t, w) \partial_w$.\\
$\bullet$~ We calculate $I(t, x, \dot x) =G\big(t, w(t, x, \dot x)\big)$.

\begin{figure}[h]
	\centering
	\includegraphics[width=0.55\textwidth]{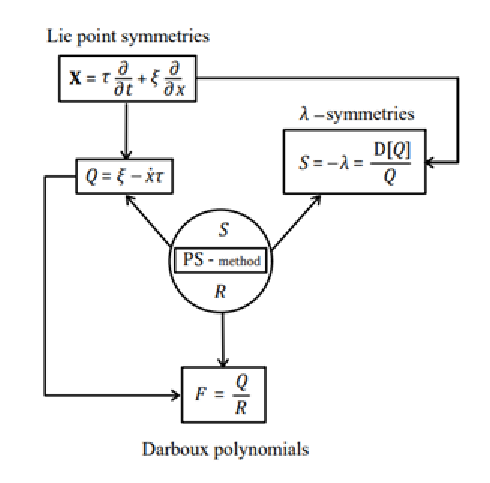}
	\caption{ Flowchart indicates the interconnections between the quantities $ \tau $, $ \xi $, $ \lambda $, $ F $, $ R $, $ S $ of the Extended PS method, $ \lambda $-symmetry,
Lie point symmetry, and DPs in the ISG-method.}
	\label{fig:mesh1}
\end{figure}

If the given DE admits a Lie point symmetry, then the $\lambda$-symmetry can be calculated
without solving the $\lambda$-prolongation condition from the relation $\lambda=D[Q]/Q$ where $D$
is the total derivative operator with respect to $t$. One quickly show that $-D[Q]/Q$ is a solution of \eqref{31}.
As the same way, the determining equation for DP associated to $ \ddot{x} =\phi(x, \dot x)$ is $D[F]=\phi_{_{\dot x}} F$.
It can be shown that $Q/F$ is also a solution of \eqref{32}.
Therefore, the characteristic $Q=\xi-\dot x \tau$ obtained from the $\lambda$-symmetry associated to the DE,
$ \ddot{x} =\phi(t, x, \dot x)$ and the DP $F$ associated to the same DE
give the integrating factor $R$ as $R=Q/F$.
Thus, by using $\lambda$-symmetry and DP without solving the determining equations,
we can obtain the $(S,R)$ as follows:
\begin{eqnarray}\label{41}
S=-\frac{D[Q]}{Q}, ~~~~~R= \frac{Q}{F}.
\end{eqnarray}
In the ISG-method, the basic quantities of the Lie point symmetry, extended PS method,  $\lambda$-symmetry and
DP are read off to be $\tau, \xi, \lambda, F, R$ and $S$. The relation between these quantities to each other is depicted as a flowchart in Figure \ref{fig:mesh1}.

\section{Friedmann equations in the presence of $\Lambda$}
\label{Sec.IV}

Our goal in this section is not actually to solve the Friedmann equations in the framework of the GR, because these
equations has already been solved in cosmology in different methods, and
their solutions can be found in textbooks such as \cite{{stephani},{Griffiths}}.
Here, we employ the ISG-method introduced in Sec. \ref{Sec.III} for solving DEs similar to
these equations in cosmology and gravity. The
reason that caused us to abandon the use of a symmetry approach alone,
such as Lie point symmetry, Noether symmetry, Hojman symmetry, etc.,
and turn to the ISG-method, is the existence of several basic problems that usually occur in
solving the DEs such as equation \eqref{2} with each of the symmetrical
methods alone. These problems implicitly will be discussed in the next section.

In order to study the FLRW cosmological model, let us consider the homogenous and isotropic universe. With a good approximation, this universe can be described
by the FLRW metric. The line element on the arbitrary point of the FLRW space-time manifold
is given by
\begin{eqnarray}\label{42}
ds^2  &=& g_{\mu \nu}~dx^\mu~ dx^\nu,\nonumber\\
&=& -c^2 dt^2+a^2(t)\Big(\frac{dr^2}{1-k r^2} + r^2 d\theta^2 + r^2 \sin^2(\theta) d\varphi^2 \Big),
\end{eqnarray}
where $x^\mu =(t, r, \theta, \varphi)$ is the space-time coordinates of the manifold centered
at any point of this universe, $a(t)$ is a differentiable function of the cosmic time $t$
which is called the scale factor, $k =0, \pm 1$ is the spatial curvature of the universe,
and $g_{\mu\nu}$'s $(\mu, \nu=0,1,2,3)$ are the components of the FLRW metric tensor $\bf g$.
In this work, our metric signature is $( - , + , + , + )$ and we
will use the natural units where the velocity of light in vacuum is
unity, $c = 1$.

The dynamics of the universe is determined by the metric tensor.
In the four-dimensional space-time, the metric tensor has ten independent components.
But the symmetries of the FLRW universe (homogeneity and isotropy) have reduced the number of the independent components from ten to one,
a single function of cosmic time, the cosmological scale factor $a(t)$. Therefore, to determine the dynamics of
the universe it is enough to find only the function  $a(t)$.
Our goal in this study is to find this function. For this purpose, we use the Einstein's theory of general relativity (GR) as the background theory.
In the theory of GR, the governing equations are the Einstein's field equations (EFEs). The EFEs in the presence of the cosmological constant $\Lambda$ are
\begin{eqnarray}\label{43}
R_{\mu \nu}+ (\Lambda-\frac{1}{2} R) g_{\mu \nu} = \kappa T_{\mu \nu},
\end{eqnarray}
where $R_{\mu \nu}$  and $R$ are the respective Ricci tensor and curvature scalar,
$T_{\mu \nu}$ is the energy-momentum tensor, and $\kappa ={8 \pi G}/ {c^4}$ in which $G$
is the gravitational constant\footnote{From now on, we use the system of units where the speed of light in vacuum is equal to 1.}.
 By defining  $p^\ast= p-\Lambda/{8\pi G}$, and $\rho^\ast =  \rho+ \Lambda/{8\pi G}$
respectively as the effective pressure and the effective density of the universe, the EFEs \eqref{43} can be written in the normal form:
\begin{eqnarray}\label{44}
G_{\mu \nu}= \kappa {T^\ast}_{\mu \nu},
\end{eqnarray}
where
\begin{eqnarray}\label{45}
G_{\mu \nu}:= R_{\mu \nu}-  \frac{1}{2} R g_{\mu \nu},
\end{eqnarray}
is the Einstein tensor, and
\begin{eqnarray}\label{46}
{T}_{\mu\nu}^\ast = \kappa  \Big(T_{\mu \nu} -\frac{\Lambda }{\kappa }g_{\mu \nu}\Big),
\end{eqnarray}
is the effective energy-momentum tensor of the matter sources in the FLRW universe.
A general form of the energy-momentum tensor of a perfect fluid by its energy density $\rho^\ast$ and the pressure $p^\ast$ is determined by
\begin{eqnarray}\label{47}
 {T}_{\mu \nu}^\ast = p^\ast g_{\mu \nu} +(p^\ast + \rho^\ast)  U_\mu U\nu.
\end{eqnarray}
Here $U^\mu = g^{\mu \nu} U_\nu$ are the contravariant components of the four-velocity vector ${\bf U}=U^\mu \partial_\mu =(1,0,0,0)= \partial_t $ of an observer comoving with the fluid.
Let us suppose the vector ${\bf U}$ has a spatial component, then the fluid has a special direction compared to the spacelike hyper-surfaces.
This would violate our assumption of spatial isotropy. On the other hand, the density and the pressure should be independent on the spatial hyper-surfaces.
Hence, they can be the function of cosmic time only, that is, $\rho=\rho(t)$ and $p=p(t)$. These functions are related by an equation of state $p=p(\rho)$.
For barotropic fluids it is given by the following linear relation
\begin{eqnarray}\label{48}
 p=  w \rho,
\end{eqnarray}
where $-1\leq w \leq 1$ is a dimensionless constant, which is called the parameter of the equation of state.
This equation covers the very important cases in cosmology, for example dust $(w=0)$, radiation $(w=1/3)$, vacuum $(w=-1)$, and stiff fluid $(w=1)$.

For the FLRW universe, we find that the off-diagonal components of the Einstein tensor are zero: $G_{\mu\nu}= 0, \mu \neq \nu$,
while the diagonal components are given by
\begin{eqnarray}\label{49}
G_{00}&=&3 \frac{{\dot a}^2 +k}{a^2} \nonumber \\
G_{ii}&=& -\Big(2 \frac{\ddot a}{a} +\frac{{\dot a}^2 +k}{a^2}\Big) g_{ii},~~i=1, 2, 3.
\end{eqnarray}
The non-vanishing components of the effective energy-momentum tensor are
\begin{eqnarray}\label{50}
{T}_{00}^\ast&=&\rho +\frac{\Lambda}{{8 \pi G}} \nonumber \\
{T}_{ii}^\ast&=& \Big(p-\frac{\Lambda}{{8 \pi G}}\Big) g_{ii},~~i=1, 2, 3.
\end{eqnarray}
Substituting the components of the tensors $G_{\mu\nu}$ and ${T^\ast}_{\mu\nu}$ respectively from Eqs. \eqref{49} and \eqref{50}
into \eqref{44} gives us
\begin{eqnarray}
3 \frac{{\dot a}^2 +k}{a^2}&=&{8 \pi G} \rho + \Lambda  \label{51} \\
-2 \frac{\ddot a}{a} -\frac{{\dot a}^2 +k}{a^2}&=& {8 \pi G} p - \Lambda.\label{52}
\end{eqnarray}
These differential equations for the unknown functions $a(t)$, $p(t)$, and $\rho(t)$ are called the Friedmann equations.
These equations form a system of two differential equations of the second-order. In order to determine unknowns uniquely,
we need also another relation between these unknowns. Here, we choose the equation of state of the barotropic fluid \eqref{48} as a new relation.
In this way, one can have a system of three differential-algebraic equations.
To solve this set of three equations, first by multiplying both sides of Eq. \eqref{51}  by $w$ one obtains
\begin{eqnarray}\label{53}
{8 \pi G} \rho w= 3w \frac{{\dot a}^2 +k}{a^2} - \Lambda w,
\end{eqnarray}
on the other hand, by plugging the equation of state \eqref{48} into Eq. \eqref{52} we arrive at the following equation
\begin{eqnarray}\label{54}
{8 \pi G} \rho w= -2 \frac{\ddot a}{a} -\frac{{\dot a}^2 +k}{a^2}+ \Lambda.
\end{eqnarray}
By combining Eqs. \eqref{53}  and \eqref{54} it is then concluded that
\begin{eqnarray}\label{55}
{\ddot a} = \alpha \frac{{\dot a}^2}{a}+ \beta a +\frac{\gamma}{a},
\end{eqnarray}
where the new parameters $\alpha, \beta$ and $\gamma$ are defined in terms of the given old parameters $w$, $\Lambda$, and $k$ as follows:
\begin{eqnarray}\label{56}
\alpha:=- \frac{1+3 w }{2},~~~~ \beta:= \Big(\frac{1+ w }{2}\Big)\Lambda,~~~~{\gamma}:= -\Big(\frac{1+ 3w }{2}\Big)k.
\end{eqnarray}
As we have seen, the combination of the set of differential-algebraic equations \eqref{51}, \eqref{52} together with \eqref{48} leads to the ODE \eqref{55}.
Indeed, Eq. \eqref{55} is a non-linear second-order ODE in which the scale factor $a(t)$ depends on the cosmic time.
This differential equation can be viewed as the equation of motion of a particle with a unit mass in one dimension.
By defining the scale factor as the coordinate of this particle, i.e., $x(t) := a(t)$,  Eq. \eqref{55} can be written in the familiar form of Newton's second law, $\ddot x = \phi(t, x, \dot x)$, where
\begin{eqnarray}\label{57}
\phi(t, x, \dot x) =\alpha \frac{{\dot x}^2}{x}+\beta x +\frac{\gamma}{x},
\end{eqnarray}
is the component of the force acting on the particle and $\ddot x$ is its acceleration. Therefore,
FLRW cosmological model in the framework of the GR theory can be imagined as the dynamics of a particle in one dimension, which is subjected to the force
\begin{eqnarray}\label{58}
{\bf F}=\phi(t, x, \dot x) \frac{\partial }{\partial x},
\end{eqnarray}
and the particle under the influence of this force has acceleration  $\ddot x$.
It is clear that for the spatially flat universe the parameter $\gamma$ is equal
to zero, and thus the equation of the particle takes the following form
\begin{eqnarray}\label{59}
\ddot x=\phi(t, x, \dot x),
\end{eqnarray}
where $\phi(t, x, \dot x)=\alpha {\dot x}^2/x +\beta x$.
Therefore, to analyze the dynamics of the spatially flat FLRW cosmological model, it is enough to solve only the DE \eqref{59}.
Our goal in the next section is to analytically solve the DE \eqref{59} in the presence of $\Lambda$ by the ISG-method.


\section{ FLRW solution in the presence of $\Lambda$ by ISG-method}
\label{Sec.V}

In order to solve the DE \eqref{59} by ISG-method, we define an one-parameter Lie group of point transformations in the one-dimensional mini-super space as in \eqref{34}.
Also, the infinitesimal generator corresponding to the infinitesimals ${\tau}(t, x)$ and ${\xi}(t, {x})$ of the group
$G$ is considered to be  ${\bf X}= \tau(t, x) {\partial_t}+\xi(t, x) {\partial_x}$.
Now, one question arises: ``what is the Lie group of point transformations \eqref{34},
so that the DE $H(t, x, {\dot x}, {\ddot x}) := {\ddot x} - \phi(t, x, {\dot x})=0$ remains invariant under that group?''.
To answer to this question, one may use Theorem 1. According to Theorem 1, the DE $H(t, x, {\dot x}, {\ddot x})=0$
remains invariant under Lie group of transformations of \eqref{34}, if and only if the Lie's invariance condition
${{{{\bf X}^{(2)}} H}{|}}_{H=0}=0$ holds, where the vector field ${\bf X}^{(2)}$
is the two-order prolongation of the infinitesimal generator $\bf X$ which is defined as follows:
\begin{eqnarray}\label{64}
{\bf X}^{(2)}= \tau(t, x) \frac{\partial}{\partial t}+\xi(t, x) \frac{\partial}{\partial x}
+\xi^{(1)}(t, x, \dot x) \frac{\partial}{\partial {\dot x}}+\xi^{(2)}(t, x, \dot x, \ddot x) \frac{\partial}{\partial {\ddot x}},
\end{eqnarray}
where the functions $\xi^{(1)}$ and $\xi^{(2)}$ are
\begin{eqnarray}
\xi^{(1)}(t, x, \dot x)&:=& \frac{d \xi^{(1-1)}}{dt} -\dot x \frac{d \tau}{dt}={\dot \xi}-{\dot x} {\dot \tau}, \label{65}\\
\xi^{(2)}(t, x, \dot x,  \ddot x)&:=& \frac{d \xi^{(2-1)}}{dt} -\ddot x \frac{d \tau}{dt}=\frac{d}{dt}({\dot \xi}-{\dot x} {\dot \tau})
-{\ddot x} {\dot \tau}={\ddot \xi}-2{\ddot x} {\dot \tau}-{\dot x} {\ddot \tau}.\label{66}
\end{eqnarray}
In order to preserve the invariance of the DE $H(t, x, {\dot x}, {\ddot x})=0$ as
\begin{eqnarray}\label{67}
H(t, x, {\dot x}, {\ddot x}) = H({\bar t}, {\bar x}, {\dot {\bar x}}, {\ddot {\bar x}}) =0,
\end{eqnarray}
under the the Lie group of transformations with the infinitesimal generator $\bf X$, we need to solve Lie's invariance condition which is
\begin{eqnarray}
0&=& {{{{\bf X}^{(2)}} H}{|}}_{H=0}\nonumber\\
&=& \Big[\tau \frac{\partial}{\partial \tau } +\xi \frac{\partial}{\partial x} +({\dot \xi}-{\dot x} {\dot \tau})\frac{\partial}{\partial \dot x}
+({\ddot \xi}-2{\ddot x} {\dot \tau}-{\dot x} {\ddot \tau})\frac{\partial}{\partial \ddot x} \Big] {(\ddot x-\phi){|}}_{H=0} \nonumber\\
&=&- \tau \phi_t-\xi \phi_x- ({\dot \xi}-{\dot x} {\dot \tau}) {\phi}_{\dot x}+{\ddot \xi}-2 \dot \tau \phi -\dot x \dot \tau.\label{68}
\end{eqnarray}
As was mentioned earlier, $\phi(t, x, \dot x)=\alpha {\dot x}^2 /x + \beta x$, then it is simply follows that
\begin{eqnarray} \label{69}
 \phi_t=0,~~~~~{\phi}_{x}=-\alpha\frac{ {\dot x}^2}{x^2} + \beta,~~~~~~~~~ {\phi}_{\dot x} =2\alpha \frac{\dot x}{x}.
\end{eqnarray}
Now by applying the total derivative operator with respect to the independent variable $t$ in the form
\begin{eqnarray} \label{70}
^{.} = \frac{d}{dt} =\frac{\partial}{\partial t}+ \dot x \frac{\partial}{\partial x} +\phi(t, x, \dot x) \frac{\partial}{\partial \dot x},
\end{eqnarray}
on the functions $\tau(t, x)$ and $\xi(t, x)$, gives us the following equations:
\begin{eqnarray}
\dot \xi &=& \xi_{_t} +\dot x  \xi_{_x} \nonumber\\
\ddot \xi &=& \xi_{_{tt}}+2\dot x  \xi_{_{xt}}+\dot {x}^2  \xi_{_{xx}}+\phi ~ \xi_{_x}, \nonumber\\
\dot \tau &=& \tau_{_t}+\dot{x}\tau_{_x}, \nonumber\\
\ddot \tau &=&  \tau_{_{tt}}+2\dot x \tau_{_{xt}}+\dot x^2 \tau_{_{xx}}+\phi \tau_{_x} . \nonumber
\end{eqnarray}
By substituting these equations into the Lie's invariance condition \eqref{68} we then obtain

\begin{eqnarray} \label{71}
\xi_{_{tt}}-\beta \xi +\beta x \xi_{_x}- \frac{2\beta}{x} \tau_{_t}+\dot x \big(-\frac{2\alpha}{x} \xi_{_t}+2 \xi_{_{tx}}-\frac{3\beta}{x} \tau_{_x} - \tau_{_{tt}}\big) \nonumber\\
+\dot x^2 \big(\frac{\alpha}{x^2} \xi-\frac{\alpha}{x} \xi_{_x}+\xi_{_{xx}}-2\tau_{_{tx}}\big)-\dot x^3 \big(\tau_{_{xx}}+\frac{\alpha}{x} \tau_{_x}\big) = 0.
\end{eqnarray}
The above equation is a polynomial equation of degree three in the variable $\dot x$ whose coefficients are linear homogenous in $\tau (t,x)$ and $\xi (t,x)$
and their partial derivatives up to second-order.
In order to always hold the Eq.\eqref{71}, the coefficients of different powers of $\dot x$ should be separately set to zero.
Accordingly, we obtain the following equations
\begin{eqnarray}
\dot x^0:&&~ \xi_{_{tt}}-\beta \xi +\beta x \xi_{_x}-\frac{2\beta}{x} \tau_{_t}=0,\label{72}\\
\dot x^1:&&~ \tau_{_{tt}} - \frac{2\alpha}{x} \xi_{_t}-2 \xi_{_{tx}}+\frac{3\beta}{x} \tau_{_x} =0,\label{73}\\
\dot x^2:&&~\xi_{_{xx}} -\frac{\alpha}{x}  \xi_{_x}+ \frac{\alpha}{x^2} \xi -2\tau_{_{tx}}=0,\label{74}\\
\dot x^3:&&~ \tau_{_{xx}}+\frac{\alpha}{x} \tau_{_x} = 0.\label{75}
\end{eqnarray}
This is a system of linear homogenous partial differential equations (PDEs) for the infinitesimals $\tau (t,x)$ and $\xi (t,x)$. This system of four linear PDEs defines the set of the ``determining equations'' for the infinitesimals $\tau $ and $\xi $  admitted by second-order ODE, $H(t,x,\dot x, \ddot x) = 0$, \cite{Bluman1}.
To solve this system of PDEs of the second-order let us consider the following ansatz for the functions $\tau $ and $\xi$ as polynomials of degree four in terms of the variables $t$ and $x$ \cite{Cantwell1}:
\begin{eqnarray}
\tau (t,x)&=& a_0+a_1t+a_2x+a_3x^2+a_4tx+a_5x^2+a_6t^3+a_7t^2x+a_8tx^2 \nonumber\\
&&+a_9x^3+a_{10}t^4+a_{11}t^3x+a_{12}t^2x^2+a_{13}tx^3+a_{14}x^4, \label{76}\\
\xi (t,x) &=& b_0+b_1t+b_2x+b_3x^2+b_4tx+b_5x^2+b_6t^3+b_7t^2x+b_8tx^2 \nonumber\\
&&+b_9x^3+b_{10}t^4+b_{11}t^3x+b_{12}t^2x^2+b_{13}tx^3+b_{14}x^4, \label{77}
\end{eqnarray}
where $a_i$'s and $b_i$'s, $(i=0, 1, \cdots, 14)$ are some unknown expansion coefficients. We must determine these coefficients so that the power series \eqref{76} and \eqref{77} are the solutions of the system of PDEs \eqref{72}-\eqref{75}. By substituting the power series expansions \eqref{76} and \eqref{77} into the determining equations \eqref{72}-\eqref{75} we arrive at the following system of equations
\begin{eqnarray}
\dot x^0:&&~ \sum_{j=0}^4\sum_{k=0}^4 p_{jk}^{(0)}t^j x^k=0, \label{78}\\
\dot x^1:&&~\sum_{j=0}^3\sum_{k=0}^2 p_{jk}^{(1)}t^j x^k=0,\label{79}\\
\dot x^2:&&~ \sum_{j=0}^4\sum_{k=0}^3 p_{jk}^{(2)}t^j x^k=0, \label{80}\\
\dot x^3:&&~  \sum_{j=0}^4\sum_{k=0}^3 p_{jk}^{(3)}t^j x^k = 0,\label{81}
\end{eqnarray}
where $p_{jk}^{(m)}$'s$~(m=0,1,2,3)$ are algebraic expressions in terms of the coefficients $a_i$'s and $b_i$'s.
In order to hold Eqs. \eqref{78}-\eqref{81}, the coefficients of different powers of $t^j x^k$  should be separately set to zero, i.e. the following equations hold
\begin{eqnarray}
p_{jk}^{(0)}=0,&&~ j=0,...,4;~k=0,...,4,\nonumber\\
p_{jk}^{(1)}=0,&&~ j=0,...,3;~k=0,...,2,\nonumber\\
p_{jk}^{(2)}=0,&&~ j=0,...,4;~k=0,...,3,\nonumber\\
p_{jk}^{(3)}=0,&&~ j=0,...,4;~k=0,...,3.\label{82}
\end{eqnarray}
This is a system of algebraic equations for the coefficients $a_i$'s and $b_i$'s.
Solving this system of the algebraic equations simultaneously, gives us the following solutions for the coefficients $a_i$'s and $b_i$'s:
\begin{eqnarray} \label{83}
a_1=a_2=...=a_{14}=0,~~~~ b_0=b_1=0,~~b_3=b_4=...=b_{14}=0.
\end{eqnarray}
Here the coefficients $a_0$ and $b_2$ can admit arbitrary values. Therefore, for a particle in the mini-super space, the most general solutions of the determining equations \eqref{72}-\eqref{75} are
$\tau (t,x)=a_0,~\xi (t,x)=b_2 x$.  By substituting these functions into the infinitesimal generator $\mathbf{X}$ we find that
\begin{eqnarray} \label{84}
\mathbf{X}=a_0 \frac{\partial}{\partial t}+b_2 x\frac{\partial}{\partial x},
\end{eqnarray}
which is a linear combination of the vector fields $\Bbb{\partial}_t$ and $x\Bbb{\partial}_x$. Therefore, there are two independent Lie point symmetries defined by the infinitesimal generators \cite{Sundermeyer1}
\begin{eqnarray} \label{85}
\mathbf{X_1}=\frac{\partial}{\partial t},~~~~~~\mathbf{X_2}=x\frac{\partial}{\partial x}.
\end{eqnarray}
The vector field $\mathbf{X_1}=\partial_t$ is the infinitesimal generator of group of translation along the time in the plane $(t,x)$
\begin{eqnarray} \label{86}
\mathbf{\Phi_1}:&&~\Bbb{R}^2 \rightarrow \Bbb{R}^2,\nonumber \\
&&~(t,x)\mapsto (\overline{t},\overline{x})=\mathbf{\Phi_1} (t,x; \epsilon)=(t+\epsilon ,x).
\end{eqnarray}
Now we consider the infinitesimal generator $\mathbf{X_2}=x \partial_x$.
To find the transformation group of $\mathbf{X_2}$ generator, we use the first fundamental theorem of Lie \cite{{Bluman1},{Ibragimov2}}
\begin{eqnarray} \nonumber
\frac{d}{d \epsilon}(\overline{t},\overline{x})=\mathbf{X_2}(\overline{t},\overline{x})=(0,\overline{x}),&~(\overline{t},\overline{x})\mid_{\epsilon =0}=(t,x).
\end{eqnarray}
Integrating in both sides of the above equations  we get $\bar{t}=t , \bar{x}=e^{\epsilon}x$.
Therefore, $\mathbf{X_2}=x \partial_x$ is the infinitesimal generator of the Lie scaling group along the $x$  in the plane $(t, x)$
\begin{eqnarray}
\mathbf{\Phi_2}:&&~\Bbb{R}^2 \rightarrow \Bbb{R}^2,\nonumber \\
&&~(t,x)\mapsto (\overline{t},\overline{x})=\mathbf{\Phi_2} (t,x; \epsilon)=(t,e^{\epsilon} x).\nonumber
\end{eqnarray}
Thus, each of the Lie transformations $\mathbf{\Phi_1} (t,x; \epsilon)=(t+\epsilon ,x)$ and $\mathbf{\Phi_2} (t,x; \epsilon)=(t,e^{\epsilon} x)$ leaves the DE $H(t,x,\dot{x},\ddot{x})=0$ invariant.

Now, to find the DP associated to the DE $\ddot{x}=\Phi(x,\dot{x}) $, we use Definition 1 of Sec. \ref{Sec.III}.
Furthermore, we consider these polynomials, which are functions of the variables $t$, $x$ and $\dot{x}$, as follows \cite{Mohanasubha1}:
\begin{eqnarray} \label{89}
F(t,x,\dot{x}):=a_0 (t,x)+a_1(t,x) \dot{x}+a_2(t,x) \dot{x}^2,
\end{eqnarray}
where $a_0$, $a_1$ and $a_2$ are some unknown functions that may depend on the variables ($t, x)$.
These functions should be determined in such a way that $F$ is an eigen-function of ``the determining equation for the DP'', $D[F]=\Phi_{\dot{x}} F$.
By substituting \eqref{89} into the determining equation $D[F]=\Phi_{\dot{x}} F$ and by employing \eqref{69} one can get
\begin{eqnarray} \label{91}
(a_{0t}+\beta x a_1)+\dot{x}(a_{1t}+a_{0x}+2\beta x a_1-2 a_0 \frac{\alpha}{x}) +\dot{x}^2 (a_1 x -a_1  \frac{\alpha}{x} +a_{2t})+\dot{x}^3 a_{2x}=0,
\end{eqnarray}
where ${a_{i}}_{t}:={\partial a_{i}}/{\partial t}, {a_{i}}_{x}:={\partial a_{i}}/{\partial x}$.
Analogously, the coefficients of different powers of $\dot{x}$ should be separately set to zero. Accordingly, we obtain that
\begin{eqnarray}
\dot x^0:&&~a_{0t}+\beta x a_1 = 0, \label{92}\\
\dot x^1:&&~ a_{1t}+a_{0x}+2\beta x a_1-2 a_0 \frac{\alpha}{x}=0,\label{93}\\
\dot x^2:&&~ a_1 x -a_1  \frac{\alpha}{x} +a_{2t}=0, \label{94}\\
\dot x^3:&&~ a_{2x} = 0.\label{95}
\end{eqnarray}
It follows from the fourth equation of this system that $a_2$ is only a function of cosmic time $t$.
Solving this system of PDEs by using the series solution method one concludes that
\begin{eqnarray} \label{96}
a_{0}(t,x)&=&-\frac{1}{\sqrt{\beta}(\alpha -1)}\Big[ \beta^{\frac{3}{2}}  (c_3\cos(2\omega t)+c_2  \sin(2\omega t))x^2 \nonumber \\
&& +\beta \sqrt{\alpha -1} (-c_4 \cos (\omega t) +c_5 \sin (\omega t)) x^{1+\alpha} \nonumber \\
&&- c_6 (\alpha -1) \sqrt{\beta} x^{2 \alpha } - c_1 \beta^{\frac{3}{2}} x^2 \Big],\\
a_{1}(t,x)&=& x^{\alpha}\big[c_4 \sin (\omega t)+c_5 \cos (\omega t)-\frac{2 \sqrt{\beta} x^{-\alpha +1}}{\sqrt{\alpha -1}} \big(c_3 \sin (2\omega t)-c_2  \cos (2\omega t) \big)\big],\label{97} \\
a_{2}(t)&=& c_1+c_2 \sin (2\omega t)+c_3 \cos (2\omega t),\label{98}
\end{eqnarray}
where $c_i$'s ~$(i=1,...,6)$ and $\omega:=\sqrt{\beta (\alpha -1)}$  are some real constants.
Using these one can obtain DP associated to the DE $H(t,x,\dot{x},\ddot{x})=0$, giving us
\begin{eqnarray}
F(t,x,\dot{x}) &=& c_1\Big(\dot{x}^2\! +\! \frac{\beta x^2}{\alpha -1} \Big)+c_2 \Big[\dot{x}^2 \sin(2\omega t)\! +\! \frac{2 \sqrt{\beta} x\dot{x}\cos(2\omega t)}{\sqrt{\alpha -1}}\! -\! \frac{\beta x^2\sin(2\omega t)}{\alpha -1} \Big]\nonumber\\
&&+c_3 \Big[\dot{x}^2 \cos(2\omega t)\! -\! \frac{2 \sqrt{\beta} x\dot{x}\sin(2\omega t)}{\sqrt{\alpha -1}}\! -\! \frac{\beta x^2\cos(2\omega t)}{\alpha -1} \Big]\nonumber\\
&&+c_4 \Big[\dot{x}x^{\alpha} \sin(\omega t)\! +\! \frac{x^{1+\alpha}\sqrt{\beta}\cos(\omega t)}{\sqrt{\alpha -1}} \Big]\nonumber\\
&&+c_5 \Big[\dot{x}x^{\alpha} \cos(\omega t)\! -\! \frac{x^{1+\alpha}\sqrt{\beta}\sin(\omega t)}{\sqrt{\alpha -1}} \Big]+c_6~x^{2\alpha} \label{99}
\end{eqnarray}
By various choices of constants $c_i$ one can consider the polynomial \eqref{99} as a linear combination of the six functions.
This set of six functions is listed in the Table 1.
\begin{center}
	\scriptsize {{{\bf Table 1.}~DP associated to the DE,  \\$~~H(t, x, \dot {x}, \ddot {x}) = \ddot {x}- \alpha {\dot x}^2 / x-\beta x=0$.  }}\\
	{\scriptsize
		\renewcommand{\arraystretch}{1.5}{
			\begin{tabular}{|p{1cm}|l|} \hline \hline
				$~~n$ & ~$F_n(t, x, \dot {x})$\\ \hline
				~~1 &  ~${\dot x}^2 + \frac{\beta }{\alpha -1} x^2$ \\
				
				~~2&  ~${\dot x}^2 \sin(2\omega t)+ \frac{2 \sqrt{\beta} x \dot x  \cos (2\omega t)}{\sqrt{\alpha -1} }-\frac{\beta x^2 }{\alpha -1} \sin(2\omega t) $ \\
				
				~~3 &  ~${\dot x}^2 \cos(2\omega t)- \frac{2 \sqrt{\beta} x \dot x   \sin (2\omega t)}{\sqrt{\alpha -1} }-\frac{\beta x^2 }{\alpha -1} \cos(2\omega t) $ \\

		     	~~4 &  ~${\dot x} x^{\alpha}  \sin(\omega t)+ \frac{ {\sqrt \beta } x^{\alpha+1}}{\sqrt{\alpha -1} }\cos (\omega t) $ \\
				
				~~5 &  ~${\dot x} x^{\alpha}  \cos(\omega t)- \frac{ {\sqrt \beta } x^{\alpha+1}}{\sqrt{\alpha -1} }\sin (\omega t) $ \\

		    	~~6 &  ~$x^{2 \alpha}$ \\
				\hline
				
	\end{tabular}}}
\end{center}

Let us now turn our attention to investigate the first integral $I (t, x,\dot{x})$ associated to the Lie symmetry vectors given by \eqref{85},
in such a way that we apply DPs presented in Table 1.\\\\
$\bullet$~{\bf The first integral corresponding to the vector $\mathbf{X}_1=\partial/\partial t$ and DP $F_6$.}
We shall consider one of the Lie symmetry vectors, for example $\mathbf{X}_1=\partial/\partial t$.
According to Definition 2 ($\lambda$-symmetry),
the characteristic corresponding to this symmetry vector is $Q_1=\xi_1-\dot{x} \tau_1=-\dot{x}$.
Using this quantity and the total derivative operator with respect to cosmic time $t$, we can calculate $\lambda_1$  associated to this characteristic which is nothing but
\begin{eqnarray} \label{101}
\lambda_1=\frac{D[Q_1]}{Q_1}=\frac{D[-\dot{x}]}{-\dot{x}}=\alpha \frac{\dot{x}}{x}+\beta \frac{x}{\dot{x}}.
\end{eqnarray}
Hence, the null form associated to the generator $\mathbf{X}_1=\partial/\partial t$ becomes $S_1=-\lambda_1=-\alpha \dot{x}/x-\beta x/\dot{x}$.
Now we find the integrating factor $R_1$ associated to  $S_1$. It follows from the characteristic $Q_1=-\dot{x}$ and DP $F_6=x^{2\alpha}$ that
$R_1={Q_1}/{F_6}=-{\dot{x}}/{x^{2\alpha}}$. Then the set $(S_1,R_1)$ is read off
\begin{eqnarray} \label{102}
(S_1,R_1)=\Big(-\alpha \frac{\dot{x}}{x}-\beta \frac{x}{\dot{x}},-\frac{\dot{x}}{x^{2\alpha}} \Big) .
\end{eqnarray}
Having the set $(S_1,R_1)$, we can construct a first integral for the $H=0$. According to \eqref{30}, the elements of first integral is made by means of the set $(S_1,R_1)$, giving us
\begin{eqnarray}
r_{_{1,1}}&=&\int R_1(\phi + \dot{x}S_1)dt=0,\nonumber\\
r_{_{1,2}}&=&-\int \big( R_1S_1+\frac{\partial r_{1,1}}{\partial x})dx=\frac{1}{2}\dot{x}^2 x^{-2\alpha}-\frac{\beta}{2(1-\alpha)}x^{-2\alpha +2}.\nonumber
\end{eqnarray}
Finally, by inserting $r_{1,1}$ and $r_{1,2}$ into Eq. \eqref{29}, the  first integral is worked out:
\begin{eqnarray}
I_1(x,\dot{x})&=&r_{_{1,1}}+r_{_{1,2}}-\int \Big[R_1+\frac{\partial}{\partial \dot{x}}(r_{_{1,1}}+r_{_{1,2}})\Big]d\dot{x}\nonumber\\
&=&\frac{1}{2}\dot{x}^2 x^{-2\alpha}-\frac{\beta}{2(1-\alpha)}x^{-2\alpha+2}. \label{103}
\end{eqnarray}
$\bullet$~{\bf The first integral corresponding to the vector $\mathbf{X}_2=x \partial_x$ and DP $F_1$.}
To get another set including the null form and integrating factor such as $(S_2,R_2)$, this time we choose the symmetry vector $\mathbf{X}_2=x \partial_x$, as well as the DP $F_1$.
Accordingly, we find that
\begin{eqnarray}
&& Q_2 = \xi _2 - \dot{x} \tau _2 = x,\nonumber \\
&& \lambda _2 = \frac{D[Q_2]}{Q_2} = \frac{D[x]}{x} = \frac{\dot{x}}{x},\nonumber \\
&& S_2 = -\lambda _2 = -\frac{\dot{x}}{x},~~~ R_2 = \frac{Q_2}{F_1} = \frac{x}{\sigma x^2 + \dot{x}^2}, \nonumber
\end{eqnarray}
where $\sigma = \beta / (\alpha -1)$. Thus,
\begin{eqnarray}
(S_2,R_2)=\Big(- \frac{\dot{x}}{x},\frac{x}{\sigma x^2+\dot{x}^2} \Big). \label{104}
\end{eqnarray}
Now we calculate the first integral associated to the set $(S_2 , R_2 )$. According to Eqs.\eqref{30}, we have
\begin{eqnarray} \label{105}
r_{_{2,1}}&=&\int R_2(\phi + \dot{x}S_2)dt = \int \frac{x}{\sigma x^2+\dot{x}^2}\Big[\alpha \frac{\dot{x}^2}{x}+\beta x+\dot{x}\Big(-\frac{\dot{x}}{x}\Big)\Big]dt,\nonumber\\
&=&(\alpha -1)t.
\end{eqnarray}
To calculate $r_{_{2,2}}$, we consider the following three different cases:\\
{\bf (a)} Case $\sigma = \beta /(\alpha -1)>0 $. For this case we find that
\begin{eqnarray}
r_{_{2,2}}&=& -\int \Big[ \Big( \frac{a}{\sigma x^2 + \dot{x}^2}\Big) \Big(-\frac{\dot{x}}{x}\Big)+\frac{\partial}{\partial x}\Big( (\alpha -1)t \Big) \Big]dx \nonumber\\
&=&\frac{1}{\sqrt{\sigma}} \tan^{-1} \big(\frac{\sqrt{\sigma} x}{\dot{x}}\big) \nonumber
\end{eqnarray}
and hence the corresponding first integral becomes
\begin{eqnarray}
I_2(t,x,\dot{x})=(\alpha -1)t+\frac{1}{\sqrt{\sigma}} \tan^{-1} \big(\frac{\sqrt{\sigma} x}{\dot{x}}\big). \label{109}
\end{eqnarray}
{\bf (b)} Case $\sigma=\beta /(\alpha -1)=0$. In this case, one gets
\begin{eqnarray} \label{106}
r_{_{2,2}}&=&-\int \Big[R_2 S_2 + \frac{\partial r_{2,1}}{\partial x}\Big]dx = -\int \Big[\Big(\frac{x}{\dot{x}^2}\Big) \Big(-\frac{\dot{x}}{x}\Big)+\frac{\partial}{\partial x}\Big( (\alpha -1)t \Big)\Big]dx,\nonumber\\
&=&\frac{x}{\dot{x}}.
\end{eqnarray}
Then, the first integral associated to the set $(S_2,R_2 )$ is worked out
\begin{eqnarray} \label{108}
I_2(t,x,\dot{x})&=&r_{_{2,1}}+r_{_{2,2}}-\int \Big[R_2+\frac{\partial}{\partial \dot{x}}(r_{_{2,1}}+r_{_{2,2}})\Big]d\dot{x}\nonumber\\
&=&(\alpha -1)t+\frac{x}{\dot{x}}.
\end{eqnarray}
{\bf (c)} Case $\sigma = \beta / (\alpha -1)<0 $. Similarly to the previous cases, we first obtain that
 \begin{eqnarray}
r_{_{2,2}}&=&-\int \Big[ \Big( \frac{a}{\sigma x^2 + \dot{x}^2}\Big) \Big(-\frac{\dot{x}}{x}\Big)+\frac{\partial}{\partial x}\Big( (\alpha -1)t \Big) \Big]dx \nonumber\\
&=&  \frac{1}{ \sqrt{|\sigma|}} \tanh^{-1} \big(\frac{\sqrt{|\sigma|} x}{\dot{x}}\big), \label{5.42}
\end{eqnarray}
then, inserting \eqref{105} and \eqref{5.42} into \eqref{29}, the first integral associated to the set $(S_2,R_2 )$ is obtained to be of the form
\begin{eqnarray}
I_2(t,x,\dot{x})=(\alpha -1)t+\frac{1}{ \sqrt{|\sigma|}} \tanh^{-1} \big(\frac{\sqrt{|\sigma|} x}{\dot{x}}\big). \label{110}
\end{eqnarray}
In summary, we found the first integrals $I_1 (x,\dot{x})$ and $I_2 (t,x,\dot{x})$ which respectively associated to the infinitesimal generators $\mathbf{X}_1=\partial_t $ and $\mathbf{X}_2=x \partial_x $.
The results obtained are summarized in the following formulae:
\begin{eqnarray}
I_1(x,\dot{x})&=&\dfrac{1}{2}\dot{x}^2 x^{-2\alpha}-\dfrac{\beta}{2(1-\alpha)}x^{-2\alpha +2}. \label{111}\\
 I_2 (t,x,\dot{x})&=&
\begin{cases}
 (\alpha -1)t~+~\frac{1}{\sqrt{\sigma}} \tan^{-1} \big(\frac{\sqrt{\sigma} x}{\dot{x}}\big) & \sigma >0, \label{112}\\
 (\alpha -1)t + \dfrac{x}{\dot{x}} & \sigma =0, \label{113}\\
 (\alpha -1)t+\frac{1}{ \sqrt{|\sigma|}} \tanh^{-1} \big(\frac{\sqrt{|\sigma|} x}{\dot{x}}\big)& \sigma <0. \label{114}
\end{cases}
\end{eqnarray}
Similarly, one can define two null forms $S_i (t,x,\dot{x}):=-D[Q_i ]/Q_i ,~ i=1,2$ and twelve integrating factors $R_{ij} (t,x,\dot{x}) :=  Q_i/F_j ,~i=1,2 ;~j=1, \cdots,6$
by making use of the the characteristics $Q_1 (t,x,\dot{x})=-\dot{x}$ and $Q_2 (t,x,\dot{x})=x$ together with DPs $F_1, \cdots, F_6$.
Therefore, a set of twelve members with these null forms and integrating factors can be constructed, which we called the PS set:
\begin{eqnarray} \label{115}
S_{_{\rm PS}}=\big\{\big(S_i , R_{_{ij}}\big)\big\}_{_{i=1,2;~j=1, \cdots, 6}}.
\end{eqnarray}
Following \eqref{29}, for each member of this set, a first integral in the form
\begin{eqnarray} \label{116}
I_{_{i,ij}}=r_{_{1;ij}}+r_{_{2;ij}}-\int \Big[R_{_{ij}}+\dfrac{\partial}{\partial \dot{x}}(r_{_{1;ij}}+r_{_{2;ij}})\Big]d\dot{x}=c_{_{i,ij}},~~~i=1, 2;~j=1, \cdots, 6,
\end{eqnarray}
can be constructed in which $c_{i,ij}$'s are some constants on the solutions of the DE \eqref{59}. Moreover, $r_{_{1;ij}}$ and $r_{_{2;ij}}$  are defined as follows:
\begin{eqnarray} \label{117}
r_{_{1;ij}}:=\int R_{ij}(\phi + \dot{x}S_i)dt,~~~~~~~~r_{_{2;ij}}:= -\int \Big[R_{ij}S_i+\dfrac{\partial}{\partial x} (r_{_{1;ij}})\Big]dx.
\end{eqnarray}
Note that all these first integrals for the DE \eqref{59} are, under the Lie point transformations, invariant.
For the sake of clarity, the members of $S_{_{\rm PS}}$ and their corresponding first integrals are listed in Table 2.
Among the first integrals in this Table 2, only two of them, $I_{1,16}$ and $I_{2,21}$, which we have denoted as $I_1$ and $I_2$, respectively, are independent and the rest are dependent on two first integrals $I_1$ and $I_2$.  Accordingly, sets $(S_1,R_{16} )$ and $(S_2,R_{21} )$ are actually the sets $(S_1,R_1 )$, $(S_2,R_2 )$, respectively,
\begin{eqnarray} \label{118}
(S_1,R_{16}):=(S_1,R_1)~~~~~ \longleftrightarrow ~~~~~ I_{1,16}=I_1,\nonumber\\
(S_2,R_{21}):=(S_2,R_2)~~~~~ \longleftrightarrow ~~~~~ I_{2,21}=I_2.
\end{eqnarray}
\begin{center}
	\scriptsize {{{\bf Table 2.}~The members of the PS set and the first integrals \\$~~~~~~~~~~~~~$associated
	to these members for the DE  $\ddot{x} =\phi (t, x, \dot x)$.  }}\\
	{\scriptsize
		\renewcommand{\arraystretch}{1.5}{
			\begin{tabular}{|p{5cm}|l|} \hline \hline
				$(S_i , R_{ij})$ &~~ ~ $I_{_{i, ij}}$\\ \hline
				~~$(S_1 , R_{11})$ &  ~~~ $I_{_{1, 11}}$ \\
				
				~~$(S_1 , R_{12})$ & ~~ ~ $I_{_{1, 12}}$ \\
				
			 ~~$(S_1 , R_{13})$ &  ~~~ $I_{_{1, 13}}$ \\

              ~~$(S_1 , R_{14})$ &  ~~~ $I_{_{1, 14}}$ \\

              ~~$(S_1 , R_{15})$ &  ~~~ $I_{_{1, 15}}$ \\

              ~~$(S_1 , R_{16}):= (S_1 , R_{1})$ &  ~~~ $I_{_{1, 16}}:=I_{_1}=E$ \\

               ~~$(S_2 , R_{21}):= (S_2 , R_{2})$ &  ~~~ $I_{_{2, 21}}:=I_{_2}$ \\

               ~~$(S_2 , R_{22})$ &  ~~~ $I_{_{2, 22}}$ \\

                 ~~$(S_2 , R_{23})$ &  ~~~ $I_{_{2, 23}}$ \\

                   ~~$(S_2 , R_{24})$ &  ~~~ $I_{_{2, 24}}$ \\

                   ~~$(S_2 , R_{25})$ &  ~~~ $I_{_{2, 25}}$ \\

                   ~~$(S_2 , R_{26})$ &  ~~~ $I_{_{2, 26}}$ \\
				\hline
				
	\end{tabular}}}
\end{center}
It should be noted that $S_1$ is the null form associated to the one-parameter Lie group of point transformations \eqref{86},
which is the translation group along time in plane $(t, x)$. As we have shown before, the infinitesimal generator of the translation group is  $\mathbf{X}_1=\partial _t$.
Therefore, the first integral associated to this symmetry vector gives the energy of the DS $S_1^2$. So, the invariant $I_1:=E$ indicates the energy of the DS.

To continue, we use the independent first integrals above to complete the problem solving. For this purpose,
we look at  two independent invariants $I_1$ and $I_2$ as a system of algebraic equations.
To solve this system of equations, i.e. to find $x$, we must remove the variable $\dot{x}$ among the equations of the system. We find $\dot{x}$ from one of the equations, for example, from $I_2 (t,x,\dot{x})=c_2$, and then put it into the other equation, $I_1 (x,\dot{x})=c_1$. The calculations are performed for three different cases $\sigma >0$, $\sigma =0$ and $\sigma <0 $ separately as follows:\\
{\bf (a)} Case $\sigma >0$. In this case, the independent invariants are as follows:
\begin{eqnarray}
I_1(x,\dot{x})&=&\dfrac{1}{2}\dot{x}^2 x^{-2\alpha}+\dfrac{1}{2} \sigma x^{-2\alpha +2}=c_1, \label{119}\\
I_2(t,x,\dot{x})&=&(\alpha -1)t~+~\frac{1}{\sqrt{\sigma}} \tan^{-1} \big(\frac{\sqrt{\sigma} x}{\dot{x}}\big)=c_2, \label{120}
\end{eqnarray}
for some constants  $c_1$ and $c_2$. Solving the invariant $I_2 (t,x,\dot{x})=c_2$ as an algebraic equation for the variable $\dot{x}$ we get
\begin{eqnarray}
\dot{x}&=&  \dfrac{\sqrt{\sigma}x}{\tan (\theta)}. \label{121}
\end{eqnarray}
where $\theta :=\sigma^{1/2} \big[c_2-(\alpha -1)t\big]$. By plugging \eqref{121} into the invariant $I_1 (x,\dot{x})=c_1 $ one can obtain
\begin{eqnarray}
\dfrac{\sigma x^{-2\alpha+2}}{2\sin^2 (\theta)} =c_1.\label{122}
\end{eqnarray}
Then, from \eqref{122} we get
\begin{eqnarray} \label{123}
x(t)=\Big(\dfrac{2c_1}{\sigma} \Big)^{^{{\dfrac{1}{2(1-\alpha)}}}}\sin ^{^{\dfrac{1}{1-\alpha}}}(\theta).
\end{eqnarray}
On the other hand, according to \eqref{56} one may write
\begin{eqnarray}
\dfrac{1}{1-\alpha}=\dfrac{2}{3(w+1)},~~~~~~~ \sigma = \dfrac{\beta}{\alpha -1}=-\dfrac{\Lambda}{3},\label{123.1}
\end{eqnarray}
Since we dealing with the case $\sigma >0$, from the second relation of \eqref{123.1} we conclude that $ \Lambda <0$.  Finally, by using \eqref{123.1} and  the definition of $\theta$ which was mentioned earlier,
one can obtain  the scale factor for the case $\sigma >0$, giving
\begin{eqnarray} \label{123.2}
x(t)=\Big(\dfrac{-6c_1}{\Lambda} \Big)^{^{{\dfrac{1}{3(w+1)}}}}\sin ^{^{\dfrac{2}{3(w+1)}}}\Big[\dfrac{3}{2} \sqrt{\frac{-\Lambda}{3}} (w+1) (t+c^{\prime}_2)\Big],~~~~\Lambda <0.
\end{eqnarray}
Here we have defined $c^{\prime}_2:=c_2/(1-\alpha)  $.\\
{\bf (b)} Case $\sigma =0$. For this case, it simply follows from the second relation of \eqref{123.1} that $\Lambda =0$.
Imposing $\sigma =0$ on the independent invariants \eqref{119} and \eqref{120} we get
\begin{eqnarray}
I_1(x,\dot{x})=\dfrac{1}{2}\dot{x}^2 x^{-2\alpha}=c_1,~~~~~I_2(t,x,\dot{x})=(\alpha -1)t+\dfrac{x}{\dot{x}}=c_2.\nonumber
\end{eqnarray}
Similarly, solving the $I_2 (t,x,\dot{x})=c_2$ for the $\dot{x} $ one obtains $\dot{x}=x/[c_2-(\alpha -1)t]$.
Then, substituting this into the $I_1 (x,\dot{x})=c_1$ we arrive at equation $x^{2(1-\alpha )} =2c_1 [c_2-(\alpha -1)t]^2$.
Finally, from the resulting equation, one obtains the scale factor in the FLRW cosmological model in the following form
\begin{eqnarray} \label{124}
x(t)=\Big[\dfrac{9c_1}{2}(w+1)^2 \Big]^{^{\dfrac{1}{3(w+1)}}}\Big(t+c^{\prime}_2\Big)^{^{\dfrac{2}{3(w+1)}}}, ~~~~ \Lambda =0.
\end{eqnarray}
{\bf (c)} Case $\sigma <0$. In this case the cosmological constant is positive, $\Lambda >0$, and the desired invariants are as follows:
\begin{eqnarray}
I_1(x,\dot{x})&=&\dfrac{1}{2}\dot{x}^2 x^{-2\alpha}-\dfrac{1}{2} |\sigma |x^{-2\alpha +2}=c_1, \label{125}\\
I_2(t,x,\dot{x})&=&(\alpha -1)t+\dfrac{1}{\sqrt{|\sigma |} } \tanh^{-1} \dfrac{\sqrt{|\sigma |} x}{\dot{x}}=c_2. \label{126}
\end{eqnarray}
From \eqref{126} it simply follows that
\begin{eqnarray} \label{127}
\dot{x}=\frac{\sqrt{|\sigma|}x}{\tanh\Big[\sqrt{|\sigma|} \big(c_2-(\alpha -1)t\big)\Big]}.
\end{eqnarray}
Inserting \eqref{127} into \eqref{125}, we get the following solution for the scale factor of the FLRW cosmological model in the case $\Lambda >0$
\begin{eqnarray} \label{128}
x(t)=\Big(\dfrac{6c_1}{\Lambda} \Big)^{^{{\dfrac{1}{3(w+1)}}}}\sinh ^{^{\dfrac{2}{3(w+1)}}}\Big[\dfrac{3}{2}\sqrt{\frac{\Lambda}{3}} (w+1)(t+c^{\prime}_2)\Big],~~~~\Lambda >0.
\end{eqnarray}
Finally, one can use the change of variable $x(t) = a(t)$ in all three cases of the solutions \eqref{123.2}, \eqref{124}  and \eqref{128} to rewrite
the scale factor of the FLRW cosmological model.
Imposing the cosmological initial condition $a(t=0)=0$ on the solutions one concludes that the integration constant $c^{\prime}_2$ vanishes.
Therefore, the cosmic scale factor $a(t)$ takes the following forms
\begin{eqnarray}\label{130}
a(t) =
\begin{cases}
\Big(\dfrac{-6c_1}{\Lambda} \Big)^{^{{\dfrac{1}{3(w+1)}}}}\sin ^{^{\dfrac{2}{3(w+1)}}}\Big[\dfrac{3}{2}\sqrt{\frac{-\Lambda}{3}} (w+1)~t\Big],~~~~~~ &~~~~~~ \Lambda<0,~~~~~\\
\Big[\dfrac{9c_1}{2}(w+1)^2 \Big]^{^{{\dfrac{1}{3(w+1)}}}}t ^{^{\dfrac{2}{3(w+1)}}}, ~~~~~~&~~~~~~ \Lambda=0,\\
\Big(\dfrac{6c_1}{\Lambda} \Big)^{^{{\dfrac{1}{3(w+1)}}}}\sinh ^{^{\dfrac{2}{3(w+1)}}}\Big[\dfrac{3}{2} \sqrt{\frac{\Lambda}{3}} (w+1)~t\Big],~~~~~~ &~~~~~~ \Lambda>0.
\end{cases}
\end{eqnarray}
In the following,  by substituting the solution above for the cosmic scale factor into the Friedmann equation \eqref{51}, the energy density of the universe can be obtained.
We proceed to calculate the energy density for all three different cases $\Lambda <0$, $\Lambda =0$, and $\Lambda >0$. They are then read off
\begin{eqnarray}\label{138}
\rho(t)=
\begin{cases}
-\dfrac{\Lambda}{8 \pi G} \frac{1}{\sin^2 \Big[\dfrac{3}{2}\sqrt{\frac{-\Lambda}{3}} (w+1) t \Big]}, &\Lambda <0,\\
\dfrac{1}{6\pi G (w+1) t^2},~~~~~~~~~~~~~~~~~~~~~~~~~~& \Lambda =0,\\
\dfrac{\Lambda}{8 \pi G} \frac{1}{\sinh^2 \Big[\dfrac{3}{2}\sqrt{\frac{\Lambda}{3}}(w+1) t \Big]}, &\Lambda >0.
\end{cases}
\end{eqnarray}
Having the energy density function of the universe and then by using the equation of state \eqref{48} one can easily obtain the pressure of the universe, giving us
\begin{eqnarray}\label{139}
p(t)=
\begin{cases}
-\dfrac{w\Lambda}{8 \pi G} \frac{1}{\sin^2 \Big[\dfrac{3}{2}\sqrt{\frac{-\Lambda}{3}} (w+1) t \Big]}, &\Lambda <0,\\
\dfrac{w}{6\pi G (w+1) t^2},~~~~~~~~~~~~~~~~~~~~~~~~~~& \Lambda =0,\\
\dfrac{w\Lambda}{8 \pi G} \frac{1}{\sinh^2 \Big[\dfrac{3}{2}\sqrt{\frac{\Lambda}{3}}(w+1) t \Big]}, &\Lambda >0.
\end{cases}
\end{eqnarray}
The EFEs in the presence of the cosmological constant $\Lambda $ is actually equivalent to the fact that the perfect fluid constituting the universe is considered as a vacuum with the equation of state  $p=-\rho $. Accordingly, the vacuum energy density can be defined by the following relation
\begin{eqnarray} \label{140}
\rho_{_{\Lambda}}(t)=\rho_{_{\Lambda}}(t=t_0)=\dfrac{\Lambda }{8 \pi G},
\end{eqnarray}
where $t_0$ denotes the present time.
As we know in addition to the vacuum energy density, the radiation and dust contribute to the fluid energy density of the universe.
We denote to these energy densities at every moment of cosmic time $t$ by $\rho _r (t)$ and $\rho _m (t)$. Now, the density parameter corresponding to the vacuum, radiation and dust is defined as
\begin{eqnarray} \label{141}
\Omega_i (t)=\dfrac{8 \pi G}{3 H^2 (t)} \rho_i (t),
\end{eqnarray}
where  $H(t)$ defined by  $H(t) := \dot{a}(t)/a(t)$  is the Hubble parameter and index $i$ denotes $``\Lambda "$, $``r"$, and $``m"$ for the vacuum, radiation and dust, respectively.
If the majority of the constituents of the universe are only these three types of fluids, i.e.
vacuum, radiation and dust, then it can be shown that the sum of the density parameters associated to these fluids is equal to 1, namely,
\begin{eqnarray}\label{143}
1=
\begin{cases}
-\Omega_{\Lambda} (t)+\Omega_r (t)+\Omega_m (t), &\Lambda <0,\\
\Omega_r (t)+\Omega_m (t), & \Lambda =0,\\
\Omega_{\Lambda} (t)+\Omega_r (t)+\Omega_m (t), &\Lambda >0.
\end{cases}
\end{eqnarray}
It should be noted that if the universe also has a spatial curvature, i.e. $k\neq 0$, then this spatial curvature also has a contribution to the energy density of the universe.
Then , the corresponding density parameter is
\begin{eqnarray} \label{144}
\Omega_k (t)=-\dfrac{k}{ H^2 (t)a^2 (t)},
\end{eqnarray}
accordingly, equation \eqref{143} may be expressed as
\begin{eqnarray}\label{145}
\Omega =1=
\begin{cases}
-\Omega_{\Lambda} (t)+\Omega_r (t)+\Omega_m (t)+\Omega_k (t), &\Lambda <0,\\
\Omega_r (t)+\Omega_m (t)+\Omega_k (t), & \Lambda =0,\\
\Omega_{\Lambda} (t)+\Omega_r (t)+\Omega_m (t)+\Omega_k (t), &\Lambda >0,
\end{cases}
\end{eqnarray}
where $\Omega $ is called the total density parameter of the universe. We note that all these density parameters are dimensionless quantities.


\section{ The age of the universe}
\label{Sec.VI}
Now in the spatially flat $(k=0)$ FLRW universe, let us now assume that the universe is filled only one fluid with the state parameter $w$.
For this case we shall calculate the age of the universe. To this end, we consider the energy density \eqref{138}. Starting from equation \eqref{141}, for the case $\Lambda<0$, one may write
\begin{eqnarray} \label{146}
\Omega (t) = \frac{8 \pi G}{3 H^2(t)}\rho (t) &=& \dfrac{|\Lambda |}{3 H^2 (t)} \dfrac{1}{\sin^2 \Big[\dfrac{3}{2}\sqrt{\frac{|\Lambda|}{3}} (w+1) t\Big]}\nonumber\\
&=& |\Omega_{_\Lambda } (t)| \dfrac{1}{\sin^2 \Big[\dfrac{3}{2}\sqrt{\frac{|\Lambda|}{3}} (w+1) t\Big]}.
\end{eqnarray}
Solving the above equation for the cosmic time $t$ we obtain
\begin{eqnarray} \label{150}
t=\dfrac{2}{3(w+1)\sqrt{|\Lambda |/3}}\sin^{-1}\sqrt{\dfrac{|\Omega_{_{\Lambda}}(t) |}{\Omega_{w}(t)}}.
\end{eqnarray}
Putting the present time $t_{0 }$ in equation \eqref{150} one can calculate the age of the universe for the case $\Lambda <0$.
For this case, since $\Lambda$ in negative, we have $\Omega_\Lambda <0$, hence we can write $\Omega_\Lambda=-|\Omega_\Lambda|$.
Then by using the fact that $\Omega_w=1-\Omega_{_{\Lambda}}$, one gets
\begin{eqnarray} \label{153}
\dfrac{|\Omega_\Lambda|}{\Omega_w}=\dfrac{|\Omega_\Lambda|}{1+|\Omega_\Lambda|}.
\end{eqnarray}
Finally, utilizing the definition $|\Lambda|/3=H_0^2 |\Omega_{\Lambda}|$, we find the formula for the age of the universe in the form
\begin{eqnarray} \label{156}
t_0=\alpha_{_\Lambda} \dfrac{2}{3(w+1)H_0}.
\end{eqnarray}
Here we have called $\alpha_{_\Lambda}:=\dfrac{\tan^{-1} (\sqrt{|\Omega_\Lambda |})}{\sqrt{|\Omega_\Lambda |}}$,  ``the age correction factor'' of the universe. Similar to above, we can calculate the age of the universe for cases $\Lambda=0$ and $\Lambda > 0$.
Therefore, the formula for the age of the universe, which includes all three cases of $\Lambda <0$, $\Lambda =0$, and $\Lambda >0$, is
\begin{eqnarray} \label{159}
t_0= \alpha_{_\Lambda} ~{\rm t_S},
\end{eqnarray}
where
\begin{eqnarray}\label{160}
\alpha_{_\Lambda} =
\begin{cases}
\dfrac{\tan^{-1} \sqrt{|\Omega_\Lambda |}}{\sqrt{|\Omega_\Lambda |}},~~~~~~ &~~~~~~ \Lambda<0,~~~~~\\
1, ~~~~~~&~~~~~~ \Lambda=0,\\
\dfrac{\tanh^{-1} \sqrt{\Omega_\Lambda }}{\sqrt{\Omega_\Lambda }},~~~~~~ &~~~~~~ \Lambda>0,
\end{cases}
\end{eqnarray}
and ${\rm t_S} :=  2/\big[3(w+1) H_0 \big]$ is the age of the universe in the cosmology standard model where the cosmological constant is not present and the dominant matter of the universe is fluid with the state parameter $w$.

In the dominant matter universe $(w=0)$, the age of the universe in standard model is equal to ${\rm t_S}=2/(3H_0)$.
On the other hand, $H_0:= H(t=t_0 )=100h$ km s$^{-1}$ Mpc$^{-1}$, where the parameter $h$ is a dimensionless number belonging to the interval $[0.55 , 0.85]$.
Choosing $h=0.674$, that is consistent with the astronomical observations, the Hubble's constant reduces to $H_0=67.4$ km s$^{-1}$ Mpc$^{-1}=6.89 \times 10^{-19}$  yr$^{-1}$ \cite{Aghanim}.
Therefore, the age of the universe with $w=0$ is
\begin{eqnarray} \nonumber
{\rm t_S}= \dfrac{2}{3H_0} \simeq 9.69 \times 10^{9}~ {\rm yr}.
\end{eqnarray}
When the dominant matter is dust, we can calculate the age correction factor of the universe. In the case $\Lambda <0$, the density parameter associated to the dust is $\Omega_m=0.315$, then $\Omega_{_{\Lambda<0}}=-0.685$.
Substituting this value into the formula \eqref{160} we find that $\alpha_{_{\Lambda<0}}=0.835$.
In the case $\Lambda>0$, the $\Omega_{_{\Lambda>0}}=0.685$, and thus $\alpha_{_{\Lambda>0}}=1.426$.
Using these together with equation  \eqref{159} one can get the age of the universe in the presence of the cosmological constant  $\Lambda$, giving us
\begin{eqnarray}\label{161}
t_0=\alpha_{_\Lambda} {\rm t_S} \simeq
\begin{cases}
8.09\times 10^9 ~ {\rm yr},~~~~~~ &~~~~~~ \Lambda<0,~~~~~\\
9.69 \times 10^9 ~{\rm yr}, ~~~~~~&~~~~~~ \Lambda=0,\\
13.82 \times 10^9 ~{\rm yr},~~~~~~ &~~~~~~ \Lambda>0.
\end{cases}
\end{eqnarray}

Before closing this section let us discuss obtaining the constant of integration $c_1$.
In order to write the constant $c_1$ and the cosmological constant $\Lambda$ in terms of the density parameters of the constituent fluids of the universe,
one has to begin the Friedmann equation.
Then, by using the fact that the cosmological scale factor at the present time $t=t_0$  is equal to 1, i.e., $a_0 := a(t=t_0 )=1$ and also by employing equations \eqref{159} and \eqref{160},
one can determine the constant $c_1$ in the solutions given by relation \eqref{130}.
Finally, the scale factor, energy density and pressure of the universe in the spatially flat $(k=0)$ FLRW cosmological model are clearly determined.

\section{Conclusions}
\label{Sec.VII}
According to contents mentioned in this work, it can be said that the
ISG-method will be a useful method for solving the DEs of
the cosmological models. This combined method has the following advantages
over other symmetry methods that are used to solve DEs:\\
$(1)$-The ISG-method is an algorithmic and comprehensive method for
solving non-linear DEs in physics, especially in gravity and cosmology.
This method can be used to solve the DEs of any cosmological model
whose force functions in the mini-super space are fractional functions of polynomials as
\begin{eqnarray}
\phi(t, x, \dot{x}) = \frac{P(t, x, \dot{x}) }{Q(t, x, \dot{x}) },
\end{eqnarray}
where $P$ and $Q$ are polynomials in the variables $t, x,$ and $\dot{x}$ with coefficients in the field of complex numbers.\\
$(2)$-If the condition $(1)$ is met, then the calculation of two independent invariants $I_1 (x, \dot{x})=c_1, I_2 (t, x, \dot{x})=c_2$ is straightforward, that is, they can be calculated by using the Theorem 2.\\
$(3)$-Unlike symmetry methods such as Lie, Noether, Hojman, Lutzky, etc, which provide for us only one first integral to solve DEs, the ISG-method provides two independent first integrals,
which makes solving the problem simple.
In the ISG-method, the independent first integrals
$I_1 (x, \dot{x})=c_1, I_2 (t, x, \dot{x})=c_2$ can be viewed as a system of algebraic equations for the unknowns $x$ and $\dot{x}$ and they can be solved simultaneously for these unknowns, which is always possible.
\\
$(4)$-If the force function of the DE is not an explicit function of time, i.e., $\phi=\phi(x, \dot{x})$, then the
translation group in the direction of time is one of the symmetry groups of the DE,
and the infinitesimal generator vector of this group of transformations is ${\bf X} =\partial/\partial t$.
\\
$(5)$-It is difficult to calculate the basic quantities of the extended PS method, i.e., the null form $S$ and the integrating factor $R$ from their DEs.
This is the only problem in the extended PS method. In the ISG-method, this problem is solved by $\lambda$-symmetry and DPs,
so that these quantities can be calculated with the help of equation \eqref{41}.
\\
$(6)$-If the DE such as $H(t, x, \dot{x}, \ddot{x})=0$ does not have any Lie point symmetry,
which we can use to indirectly calculate the basic quantities by the extended PS method,
i.e., $(S, R)$, then one has to look at the generalized symmetries. One of these well-known symmetries is the $\lambda$-symmetry which was introduced in Definition 2 of Sec. \ref{Sec.III}.

In summary, in this work, we have provided an analytical and algorithmic method to obtain the solutions of the DEs of the spatially flat (k=0) FLRW cosmological model.
The DEs of this model in the framework of the GR theory, and many other models
in the framework of the extended gravity theories such as the Brans-Dicke theory, $f(R)$ theory, Rastall theory, and etc.
have been solved analytically and numerically.
But a solution method that is general, so that it can be used not only to solve the FLRW model but also the DEs of any cosmological model within the framework of any gravity theory, has not been presented yet.
There is no mathematical coherence and regularity in these solutions (presented so far).
For example,  to solve the DEs in one cosmological model, Lie symmetry is used, and in another model, Noether symmetry, and etc.
However, it is not possible to find a specific algorithm among these different solution methods, so that it solves the DEs of these models.
The extended PS method, which was later modified and completed by Chandrasekhar and et al.,
and then combined with other symmetries, became what is used in this work under the name of the ISG-method.
It has made great progress in the field of theory of non-linear differential equations theoretically.

Accordingly,  we have shown that the FLRW cosmological model in the framework of the GR theory can be imagined as the dynamics of a particle in one dimension, in
such a way that we have reduced the Friedmann equations to a non-linear second-order ODE.
Then we have analytically solved the Friedmann equations by ISG-method and have obtained the solutions of the spatially flat $(k=0)$ FLRW cosmological model in the presence of $\Lambda$.
Finally, as an application of the solutions we have looked at the age of universe in the presence of the cosmological constant where the dominant matter of the universe has been considered to be fluid with the state parameter $w$.
In this regard, we have calculated the age of universe when the dominant matter is dust.
It has been shown that in the presence of the cosmological constant in the EFEs, the age of the universe changes by a factor called the age correction factor $\alpha_{_\Lambda}$.
The positive cosmological constant causes the age of universe to increase from its value in the Friedmann's standard model $ {\rm t_S}  \simeq  9.69 \times 10^9 ~{\rm yr}$ to
$t_0  \simeq  13.82 \times 10^9~ {\rm yr}$, while the negative cosmological constant reduces the value of age from the standard value $ {\rm t_S}  \simeq  9.69 \times 10^9~ {\rm yr}$ to the value
$t_0  \simeq  8.09 \times 10^9~ {\rm yr}$. Since the positive cosmological constant helps to solve the problem of age,  the cosmological model with  $\Lambda >0$ can be an acceptable model.
On the contrary, cosmological model with $\Lambda <0$ is unacceptable, because the presence of cosmological term in the EFEs causes the age of universe to be less than the unacceptable standard value. Therefore,
the model with  $\Lambda <0$ does not help to solve the problem of the age of universe.


\subsection*{Acknowledgements}

This work has been supported by the research vice chancellor of Azarbaijan Shahid Madani University under research fund No. 1402/917.

\end{document}